\documentclass[11pt,english,twoside]{article}
\pdfoutput=1
\usepackage{jheppub}

\usepackage{lscape}
\usepackage{amsmath}
\usepackage{xcolor}
\usepackage{graphicx}% Include figure files
\usepackage{dcolumn}% Align table columns on decimal point
\usepackage{bm}% bold math

\def\p{\partial}

\def\th{\theta}

\def\-{\hphantom{-}}
\def\ov{\overline}
\def\s2{\frac{1}{\sqrt2}}

\def\be{\begin{equation}}
\def\ee{\end{equation}}
\def\bea{\begin{align}}
\def\eea{\end{align}}
\def\beqa{\begin{eqnarray}}
\def\eeqa{\end{eqnarray}}

\def\dth{{\dot{\tilde\theta}}}
\def\dsi{{\dot{\tilde\sigma}}}
\def\th{{\tilde \theta}}
\def\si{{\tilde \sigma}}

\def\ca{{\cal A}}

\def\cc{{\mathcal C}}

\def\cg{{\mathcal G}}

\def\cb{{\cal B}}

\def\cv{{\mathcal V}}

\def\mg{m_{3/2}}
\def\mg2{m^2_{3/2}}

\def\Dsl{\,\raise.15ex\hbox{/}\mkern-13.5mu D} %this one can be subscripted

\def\mff{\mathfrak f}

\newcommand{\eq}[1]{\begin{equation}
                     \begin{split} #1 \end{split}
                     \end{equation}}

\begin{document}

\title{Two-field axion inflation and the swampland constraint in the flux-scaling scenario }% Force line breaks with \\

\author[a]{Cesar Damian} 
\author[b]{and Oscar Loaiza-Brito}
\affiliation[a]{Departamento de Ingenier\'ia Mec\'anica, Universidad de Guanajuato, \\
Carretera Salamanca-Valle de Santiago Km 3.5+1.8 Comunidad de Palo Blanco, Salamanca, Mexico}
\affiliation[b]{Departamento de F\'isica, Universidad de Guanajuato, \\
Loma del Bosque No. 103 Col. Lomas del Campestre C.P 37150 Leon, Guanajuato, Mexico.}

\emailAdd{cesaredas@fisica.ugto.mx}
\emailAdd{oloaiza@fisica.ugto.mx}

\date{\today}% It is always \today, today,
             %  but any date may be explicitly specified

\abstract{
   Based on the flux-scaling scenario we study a model consisting on Type IIB string theory compactified on a Calabi-Yau manifold with a frozen complex structure in the presence of generic fluxes. The model contains (meta)stable Minkowski and de Sitter vacua as well as inflationary directions driven by two independent linear combination of axions. Due to a numerical control by fluxes, we show that cosmological parameters as the spectral index, tensor-to-scalar ratio and non-Gaussianities can be kept within observed bounds  while preserving the desired hierarchies on physical scales. Moreover we compute the deviation of the inflationary  trajectories from geodesics on field space in terms of the fluxes showing that for some regions,  they fulfill the recent proposed swampland criterion for multi-field scenarios.  
}
\arxivnumber{}


\maketitle

\section{Introduction}
Since its conception, the inflationary mechanism \cite{Guth:1980zm} has become one of the most promising models to describe the very early stage of the universe by generically explaining some theoretical and observational puzzles of present measurements. Additionally, observational evidence is confident with a nearly scale-invariance spectrum and a tensor to scalar ratio lower than 0.1 \cite{ade2016planck} implying a Universe with a gaussian CMB profile. 
Although single field inflation matches with several cosmological observables, multi-field inflation is not completely ruled out \cite{ade2016planck}. In particular, in single field inflation the primordial non-Gaussianities are suppressed  by powers of the slow-roll parameters \cite{allen197nongaussian,Maldacena:2002vr}.  Thus, it is expected that non-linear interaction with other fields shall produce observable non-Gaussianities within observational data \cite{Ade:2015ava}. Naturally,  non-Gaussianities are expected to be large in multi-field inflationary scenarios \cite{Salopek:1991jq,Sakharov:1993qh,Linde:1996gt,Bartolo:2001cw,Bernardeau:2002jy,Battefeld:2006sz}. It is our goal to show that non-Gaussianities can be kept under control, this is, with values below the present observational bounds, within the context of the recently proposed flux-scale  scenario  \cite{Blumenhagen:2015kja}.\\

The use of string theory as a quantum gravity theory to model an effective and consistent inflationary setup follows from the fact that inflation is very sensitive to UV corrections, mainly due to the high energy scale at which it is expected to occur.  In this context, the inflationary potential is identified with one or more directions on the scalar potential (constructed at the level of supergravity) which in turn depends on scalar fields arising from a string compactification. Such dependence is achieved  by turning on string fluxes supported on non-contractible cycles of the internal space, leading to the well--known moduli stabilization scenario. Stabilization of moduli  has opened up a window not only for effectively reproducing inflation but also for giving a consistent setup on which UV corrections of the scalar potential are under control as a consequence of shift symmetries on the ten-dimensional theory and their inheritance on the corresponding four-dimensional axions. In this way, one of the best motivated and supported models of inflation from string theory is  F-term axion monodromy inflation \cite{Marchesano:2014mla, Blumenhagen:2014nba,Kobayashi:2015aaa}. \\

Besides the above,  F-term axion monodromy inflation also allows us to expect polynomial suppressed ratios of different scales by considering a tree-level superpotential depending on all moduli,  including K\"ahler moduli,  which generically have been stabilized by non-perturbative effects \cite{Balasubramanian:2004uy,Balasubramanian:2005zx}.  The perturbative dependence on the K\"ahler moduli  is accomplished by turning on non-geometric fluxes \cite{Shelton:2005cf,Shelton:2006fd,Wecht:2007wu} whose presence in string theory is supported by recent studies in generalized geometry and double field theory \cite{Aldazabal:2013sca,Berman:2013eva,Andriot:2011uh,Geissbuhler:2013uka,Blumenhagen:2013hva}. Since their incorporation to string modeling, a lot of work has been done, proving that dS vacua are possible at tree level \cite{Font:2008vd,Guarino:2008ik,deCarlos:2009qm,Aldazabal:2011yz,Dibitetto:2011gm,Blaback:2013ht,Damian:2013dq, Damian:2013dwa,Hassler:2014mla,Blumenhagen:2015xpa,Blumenhagen:2015kja,Blaback:2015zra,Herschmann:2017ygb}. \\

Moreover, in the F-term axion inflation scenario, one expects that inflationary directions are driven by axions with a K\"ahler potential depending solely on the corresponding saxions.
This scheme guarantees that inflationary directions do not get quantum corrections from the K\"ahler potential, allowing the possible construction of inflationary models with large values of $r$. However, in order to have a physical consistent frame, it is necessary to have some hierarchies on the different scales as\footnote{A relation between mass hierarchies and the presence of inflationary directions was pointed out in Appendix A in \cite{Damian:2013dq}, while the role played by the conifold on the existence of hierarchies and inflation was studied in \cite{Bizet:2016paj,Blumenhagen:2016bfp}.}
\eq{
M_\text{s}>M_\text{KK}>M_\text{inf}>M_\text{i, mod}>H_\text{inf}>M_\theta,
}
where $\theta$ is the inflaton. The flux-scaling scenario \cite{Blumenhagen:2015kja,Font:2017tuu,Blumenhagen:2015jva,Blumenhagen:2015xpa} establishes a mechanism to accomplish all the above requirements by having a parametrical control of the different scales and mass of the moduli by the presence of different kind of fluxes which precisely generates a perturbative tree-level superpotential on moduli.  Roughly speaking, it is required a model with a symmetric superpotential on $n$ complex fields in the presence of $2n-1$ real fluxes. This leads to a scalar potential with a flat direction on a linear combination of axions, while the saxions, appearing on the K\"ahler potential are stabilized at a non-supersymmetric minimum. The flat direction is uplifted by parametrically breaking the symmetry of the superpotential on the complex moduli by adding an extra flux \cite{Blumenhagen:2014nba}. This addition brings a new term in the superpotential which describes an interaction between the above mentioned moduli. Under this scheme, it was shown that it is possible to find non-supersymmetric vacua free of tachyons by considering  $p$-fluxes,  while in \cite{Blumenhagen:2015xpa} it was explored a specific scenario in which a non-supersymmetric AdS vacuum  (with a flat direction) is uplifted to dS by turning on non-geometric fluxes which contribute to D-terms in the scalar potential.\\

Flat directions on the scalar potential driven by one or more moduli are of particular interest to construct inflationary scenarios corresponding to effective field theories with scalar fields coupled to gravity. However, it has been difficult to distinguish weather the models obtained from string theory correspond to those models belonging to consistent quantum theories of gravity at high energies. The set of those models not connected to consistent theories of quantum gravity has been called the ``string swampland". With the purpose to identify such regions it was recently proposed a criterion which seems to rule out the possibility to reproduce single-field inflation on string scenarios \cite{Obied:2018sgi, Agrawal:2018own,Heisenberg:2018yae}. This criterion was studied in the context of multi-field inflation in \cite{Achucarro:2018vey} where it was shown that it is possible to have inflation driven by two or more scalar fields while fulfilling the swampland criterion. Therefore it is interesting to study the possibility to have inflation on the scale-flux scenario  and weather this construction allows us to construct inflationary directions driven by more than two moduli while satisfying the swampland criterion on the slow-roll conditions.\\

Summarizing, we are interested in  a model in which we can have a hierarchy of physical scales under a parametrical control by fluxes, with a stable minimum and inflationary directions. A minimalistic model consisting on an effective theory with two-scalar fields seems to be a simple scenario to look for all the above conditions. In this work, by selecting a particular model presented in \cite{Blumenhagen:2015kja},  we study a concrete example in which inflationary directions are driven by two scalar fields corresponding to linear combinations of axions. Since in a multi-field scenario it is possible that non-Gaussianities are present, we show that under numerical control by fluxes, non-Gaussianities can also be bounded below observed values. Moreover, we show that for some inflationary trajectories around a Minkowski minimum, the swampland criterion is fulfilled in the context of multi-field inflation  described in \cite{Achucarro:2018vey}.  For that we follow the flux-scaling scenario by identifying the inflationary trajectories to those corresponding to unstabilized moduli which are slightly uplifted by the addition of $p$-fluxes. In the approximation of small $p$ we show that
the curvature parameter of  inflationary trajectories can be written as function of fluxes allowing us to get large values of it by a parametrical control of fluxes. The ratio between this parameter and the Hubble constant, which determines whether or not the swampland criterion is satisfied, is also larger than one on different zones of the trajectories. However, this ratio cannot be parametrically controlled by fluxes, at least for constant values. It is also important to mention that for all our analysis it is necessary to relax the integer quantization requirement on fluxes. We comment on this issue in our final remarks.\\

Our work is organized as follows: In section II we review the model studied in \cite{Blumenhagen:2015kja} and present our two-field inflationary model. We also compute the ratio of the different scales parametrically controlled by  3 different fluxes. Section III is devoted to compute cosmological parameters as the tensor to scalar ratio, mass hierarchy, the spectral index and non-Gaussianities. We show that all these values can be expressed in terms of a ratio of fluxes. However, as mentioned, integer quantization of fluxes seems to bring numerical values beyond observations. This issue must be understood and carefully studied. In section IV we explore the values of the inflationary swampland criterion for the above inflationary trajectories. Finally, we present our conclusions and an Appendix devoted to review some useful notation.

%--------------------------------------------------------------------------------------------------------
\section{String compactification with non-geometric fluxes}

We shall select a model studied in detailed in \cite{Blumenhagen:2015kja}  consisting on Type IIB superstring compactified on Calabi--Yau manifolds in the presence of orientifold 3-planes with non-vanishing fluxes, including the common NS-NS and RR fluxes from the closed string sector and the so called non-geometric fluxes $Q$ and $R$ \cite{Shelton:2005cf,Shelton:2006fd,Wecht:2007wu}. Specifically we study the case for a CY manifold with a frozen complex structure and a single K\"ahler modulus, this is $h_-^{2,1}=0$ and $h^{1,1}_+=1$. We have also limited our model to manifolds for which $h^{1,1}_-=0$ meaning that the closed string potential form called the geometric moduli $G_2=Sb_2+c_2$ is absent in the superpotential,  and with $h^{2,1}_+=1$ allowing the presence of D-terms.  For this case, the K\"ahler potential reads
\eq{
\label{KSTU}
K=-3 \log (T+\ov T)-\log (S+\ov S) \, ,
}
with a perturbative dependence on the complex moduli fields $S=e^{-\phi}+ic$ and $T=\tau +i \rho$.  The superpotential depends on the coupling between RR and NS-NS fluxes with the complex structure, the geometric moduli  with the odd K\"ahler moduli (which we consider absent) and the non-geometric fluxes $Q$ with the K\"ahler moduli. Notice that the $R$ flux is not present in the superpotential. Therefore, by turning on 3 real fluxes $f, h$ and $q$  the superpotential is 
\eq{ \label{wpot}
W = if - i hS  -i qT  \, , 
}
showing a linear dependence on $S$ and $T$, where $f, h$ and $q$ are {\it in principle} integer quantized RR, NS-NS and $Q$ fluxes respectively. The scalar potential constructed from the superpotential is given by the $N =1$ supergravity scalar potential 
\eq{
V_F= \frac{M_{Pl}^4}{4\pi} e^K \left( K^{I\bar{J}}D_IW D_{\bar{J}}\bar{W}-3|W|^2\right) \,.
}
This model presents an explicitly manifestation of the scenario proposed in \cite{Blumenhagen:2015kja} in which we have 3 fluxes and 4 real moduli. It is then expected that 3 out of 4 real moduli get a vacuum expectation value while a flat direction would be present. In fact, as shown in \cite{Blumenhagen:2015kja}, the F-term potential generated by \eqref{wpot} has an interesting non-supersymmetric and non-tachyonic AdS extremum, with a scalar potential
\eq{
V_F=\frac{M^{4}_{Pl}}{64 \pi}\left(  \frac{(hs-\mff)^2}{s\tau^3} -\frac{2q(3hs+\mff)}{s\tau^2} -\frac{5q^2}{3s\tau}+\frac{(hc+q\rho)^2}{s\tau^3}\right) \,,
}
depending solely on $\tau$, $s$ and $ \theta$ where $s=e^{-\phi}$ and
\eq{ \label{eq:ax1}
\theta = h c + q \rho \, ,
}
with the minimum at $(s, \tau, \theta)=(f/h, 6f/5q,0)$ and a value given by
\eq{
V_{0F}= - \frac{\text{M}_{\text{Pl}}^4}{2^4 \pi} \frac{5^2}{3^3} \frac{q^3 h}{f^2} \,, 
}
It is important to mention that  for $|f/h| \gg 1$ and $|f/q| \gg 1$, this models exhibits a weak string coupling and large radius implying that higher-order corrections of the scalar potential can be safely ignored.
 These fluxes are subject to the Bianchi identities and tadpole cancellation conditions given in \cite{Blumenhagen:2015kja}, however,  Bianchi identities are trivially fulfilled by the selected fluxes. Notice as well that for the fixed moduli to be realistic, the value of $V_{0F}$ must be negative exhibiting a non-SUSY AdS minimum along 3 moduli directions out of 4. Since the orthogonal direction to $\theta$ is flat what we have constructed here is not a truly non-SUSY AdS vacuum, i.e, we must not worry about its stability. However,  the stable non-supersymmetry configuration allows us to avoid the appearance of tachyons within a possible uplift to dS \cite{Conlon:2006tq}.

\subsection{Uplifting to metastable dS by D-terms}

We now want to analyze the effect of adding the D-term potential allowed for $h^{2,1}_+=1$ closely following \cite{Blumenhagen:2015xpa} to uplift to Minkowski or dS minimum (still, with a flat direction). For that it is considered the presence of extra non-geometric fluxes $f$ and $R$ which survive the orientifold projection and appear in the D-term of the scalar potential. Such a term was computed in \cite{Blumenhagen:2015xpa} and by turning on just two fluxes, $r$ and $g$ respectively, it reads
\eq{
\label{dpot3}
V_D= \frac{\delta}{v \tau^2} \left(g - \frac{ r \tau}{ 3  s}\right)^{\!\! 2} \, ,
}
where $\delta$ is a positive constant. Observe that this term depends on all
the saxions in the model. The fluxes entering in $V_D$ are related to the action of the twisted differential
${\mathcal D}$ on the even $(2,1)$ forms. Such fluxes do not enter at all in the superpotential $W$ that determines
the F-term potential. However, there are Bianchi identities that mix $r$ and $g$
with NS-NS and $Q$-fluxes appearing in $W$. In our case there is only one non trivial BI constraint given by
\eq{
\label{biadd}
r h - g q = 0  \, .
}
We find that the model admits an uplift that can be either to Minkowski or dS depending on the value of $\delta$. 
Due to the complexity of the solution, we limit ourselves to present the values of stabilized moduli  as  Taylor
series expansions to linear order in $V_0=\Lambda$. For an arbitrary value of $\Lambda$, there is still an extremum at $\theta=0$
whereas the saxions are stabilized at
\eq{
s &= \frac{1}{2^3}\frac{f}{h}-\frac{3^4}{2^7}\frac{f^3}{q^3 h^2}\Lambda + \mathcal{O}(\Lambda ^2) ,\\
\tau &= -\frac{3^2}{2^3} \frac{f}{q} + \frac{11 \cdot 3^4}{2^8} \frac{f^3}{q^4 h}\Lambda  + \mathcal{O}(\Lambda ^2)\, .
}
The constant $\delta$ in the D-term potential is given by
\eq{
\delta = -\frac{1}{2^4}\frac{q h}{g^2}+\frac{3^4}{2^{10}}\frac{f^2}{g^2 q^2}\Lambda + \mathcal{O}(\Lambda ^2)\, .
}
We then conclude that for small $\Lambda > 0$ and $h, f > 0$, we stay in the physical region if $q < 0$  which
also guarantees that $\delta > 0$. In general, to attain the perturbative regime we take $|f| > q,h$ for integer fluxes\footnote{But $|q|>f,h$ for fractional values as the case we shall consider to reproduce inflationary trajectories.}.
We remark that the above series converge for small enough $\Lambda$. Clearly for $\Lambda=0$ we have a Minkowski extremum along 3 out of 4 directions.  
Notice that the scalar potential $V=V_F+V_D$ still has a flat direction on an orthogonal direction to
$\theta$ and although we shall uplift it,  vevs and masses for all moduli are not going to be affected. \\

Hence, at the minimum the moduli acquire the following masses:
\eq{M^2_{\rm mod,i} = \left( - \mu_i \frac{q^3 h}{f^2}  + \tilde{\mu}_i \Lambda + \mathcal{O}(\Lambda ^2) \right) \frac{M^2_{\rm Pl}}{4 \pi} , \,
}
with the coefficients given by
\eq{
\mu_i = \{ \frac{2^6 (17+\sqrt{181})}{3^6}, \frac{2^6(17- \sqrt{181})}{3^6}; \frac{2^6 \cdot 7}{3^6}, 0 \}\, ,
}
and
\eq{
\tilde{\mu}_i = \{ -\frac{19 \cdot 43}{2 \cdot 3^3}-\frac{5591}{2 \cdot 3^3 \cdot 181^{1/2}}, -\frac{19 \cdot 43}{2 \cdot 3^3}+ \frac{5591}{2 \cdot 3^3 \cdot 181^{1/2}};\\ \frac{7 \cdot 19}{3^3},0 \}\, .
}
The first (last) two entries are related to linear combination of saxions (axions). \\

However the mass of moduli are altered by the value of $\Lambda$. We observe that for small enough $\Lambda$ there are no tachyons but as $\Lambda$ increases,
the first normalized mass eigenvalue becomes negative. Therefore, in order to have a stable minimum, it is required to keep a small value for $\Lambda$.\\

Finally, the string and KK mass are given by
\begin{align} \label{eq:mskk}
M^2_{\rm s} &= \frac{2^{9/2} \pi}{3^3} \frac{q^{3/2} h^{1/2}}{f^{2}} M^2_{\rm Pl} + \frac{17 \pi}{2^{3/2}} \frac{\Lambda}{q^{3/2} h^{1/2}} M^2_{\rm Pl} + \mathcal{O}(\Lambda^2 ) \, , \nonumber \\ 
M^2_{\rm KK} &= \frac{2^{4}}{3^4} \frac{q^2}{f^2} \frac{M^2_{\rm Pl}}{4 \pi} + \frac{11}{3^2} \frac{\Lambda}{q h} \frac{M_{\rm Pl}^2}{4 \pi} + \mathcal{O} ( \Lambda^2) \, ,
\end{align}
where we have used  that \cite{Buchmuller:2014vda}
\eq{ \label{eq:ms}
M^2_{\rm s} = \frac{\pi}{s^{1/2} \cv^{1/2}} \,, \quad M^2_{\text{KK}} = \frac{1}{4\pi \cv^{4/3}} \,.
}
%}
%and
%\eq{ \label{eq:mkk}
with $\cv = (2 \tau)^{3/2}$  the volume modulus of the CY in the Einstein frame and in Planck units. \\

From Eqs. (\ref{eq:mskk}) we observe that the net contribution of a positive cosmological constant diminishes the values of $M_{\rm s}^2$ and $M_{\rm KK}^2$ at sub-leading order. 
The ratio of KK to string scale at linear order in $\Lambda$ is
\eq{
\frac{M^{2}_{\rm KK}}{M^2_{\rm s}} = \frac{1}{3 \cdot 2^{5/2} \pi^2} \frac{(-q)^{1/2}}{h^{1/2}} + \frac{3 \cdot 7}{2^{17/2} \pi^2} \frac{f^2 \Lambda}{(-q)^{5/2}h^{3/2}} + \mathcal{O}(\Lambda^2) \, .
}
Thus, we have parametrical control of the KK mass over the string mass for $|q| < h$, as pointed out in \cite{Blumenhagen:2015kja} and for integer fluxes. Indeed, 
the ratio of the moduli to KK scale has the same behavior as in the non-supersymmetric non-tachyonic AdS extremum in \cite{Blumenhagen:2015kja}, which for sake of completeness we present here:
\eq{
\frac{M^{2}_{\rm mod}}{M^2_{\rm KK}} = -\frac{3^4}{2^4} h q \mu_i + \left( \frac{11 \cdot 3^6}{2^8} \frac{f^2 \mu_i}{q^2} - \frac{3^4}{2^4} \frac{f^2 \mu_i}{2^4 q^2} \right)\Lambda + \mathcal{O}(\Lambda^2) \, ,
}\\
%

%{\bf Notice that up to here we have not encountered a reason to abandon integer fluxes. As we shall see, this must be done in order to reproduce slow-roll inflation.}

\subsection{Uplifting the flat direction}

In order to unflatten the orthogonal direction to $\theta$, we follow the same receipt proposed in \cite{Blumenhagen:2015kja} by considering turning on a p-flux in a generic $\Lambda \geq 0$ vacua, with superpotential
\eq{
W = \lambda W_0 + p S T \,,
}
where $W_0$ is the original superpotential with a flat direction  and $\lambda$ is a scaling parameter. Remember that our purpose is to  slightly modify the flat direction in order to reproduce inflation. The lightest modes are two orthogonal axions, given by $\theta$ and  
\eq{
\sigma = -\frac{q}{s^2} c - 3 \frac{h}{\tau^2} \rho \,.
}
In terms of canonical terms, the corresponding normalized fields read
\eq{
\tilde \theta ^2 = \frac{3 \theta^2 }{12 h^2 s^2+ 4 q^2 \tau^2} \quad \text{and} \quad \tilde \sigma ^2 = \frac{s^2 \tau^2 \sigma^2}{12 h^2 s^2+ 4 q^2 \tau^2} \,,
}
where the saxionic fields stay fixed at the minima. This condition is expected to be valid due to the mass hierarchy.\\

Up to now,  we have assumed the presence of integer fluxes. Although it is possible to uplift the former flat direction by turning on an integer and constant $p$ flux we see that one could destabilize the vevs of the 3 other moduli. Hence in order to break this structure without destabilizing the former vacuum, {we find that we must abandon the presence of constant and integer fluxes and in particular we select a non-constant  $p$ flux varying linearly along $\widetilde{\theta}$, this is $p(\theta) = - \tilde \theta$}. To preserve parametrical control of the masses, we choose $\lambda = 1/50$. In this way there is a Minkowski or dS minimum according to the value selected for $\Lambda$, which as said, must has a small value in order to avoid tachyonic directions with no flat directions. As we shall observe, selecting a non-constant flux will also allows us to find inflationary trajectories which otherwise would be impossible.  This was also observed in \cite{Blumenhagen:2015xpa}. Therefore it seems that, at least for these models in the flux scale scenario, inflation and constant-integer fluxes are not  compatible. One could think that non-constant fluxes must be discarded but several studies have shown otherwise \cite{Candelas:2014jma, Candelas:2014kma, Damian:2016lvj}. We shall comment on this important issue later on in the context of the swampland criterion.\\

The masses now satisfy  the proposed hierarchy:
\eq{
m^2_{\rm s} >  m^2_\rho > m^2_{\tilde \sigma} \sim m^2_{\tilde \theta} \,,
}
where the orthogonal axionic masses are of the same order. Thus, it is natural to identify the two orthogonal axions as the scalar fields driving inflation. It is interesting to note that the lightest scalar is in the sgoldstino direction. \\

Since we are dealing with two scalar fields, it is necessary to look for inflationary trajectories in the context of multi-field inflation. In such a framework the slow-roll parameter for single field inflation are generalized according to \cite{Gao:2014fva}  (see Appendix A for specifics in our model).\\
\begin{figure}[htbp]
   \centering
   a) \includegraphics[width=7cm]{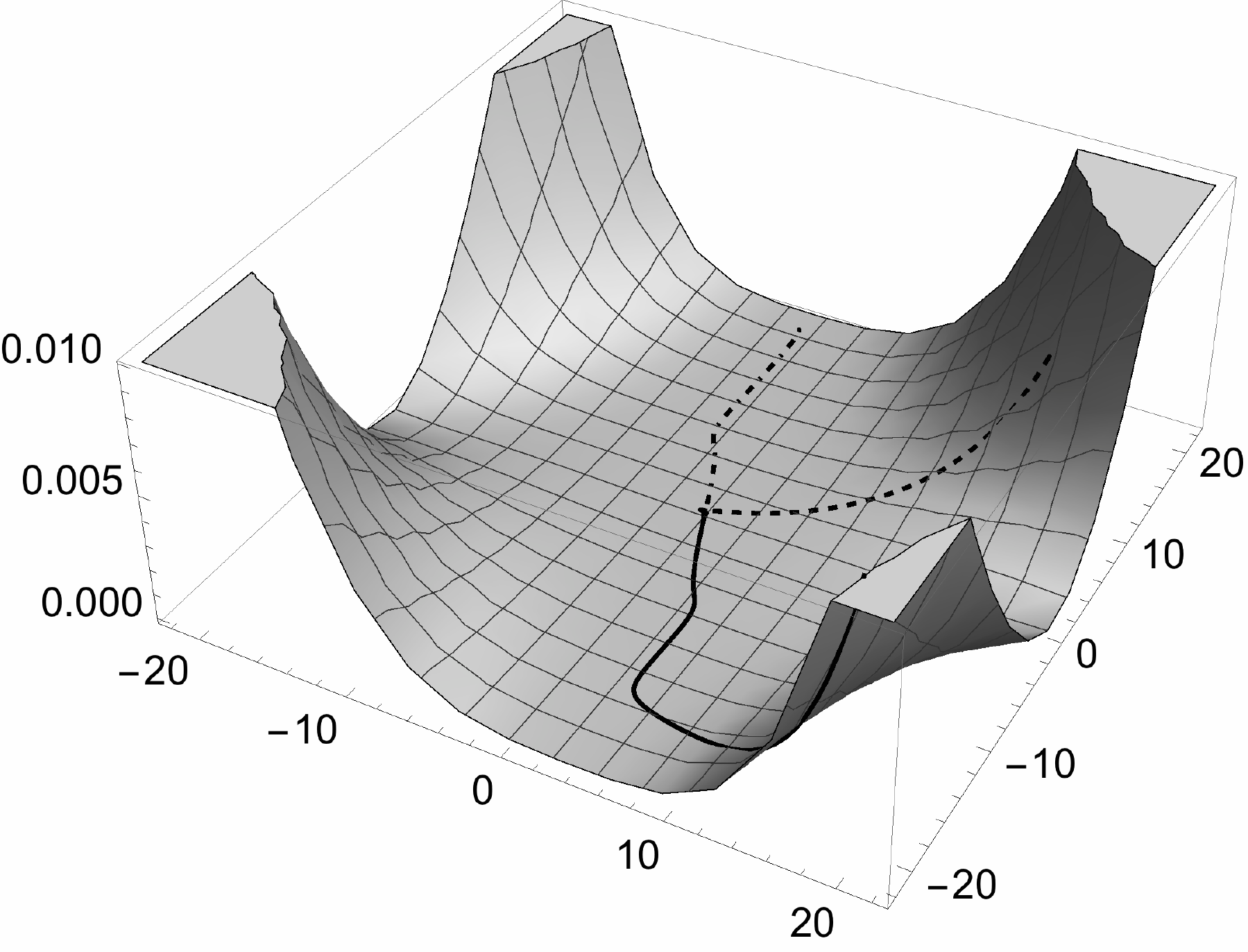} \\ \vspace{1cm}
   b) \includegraphics[width=6cm]{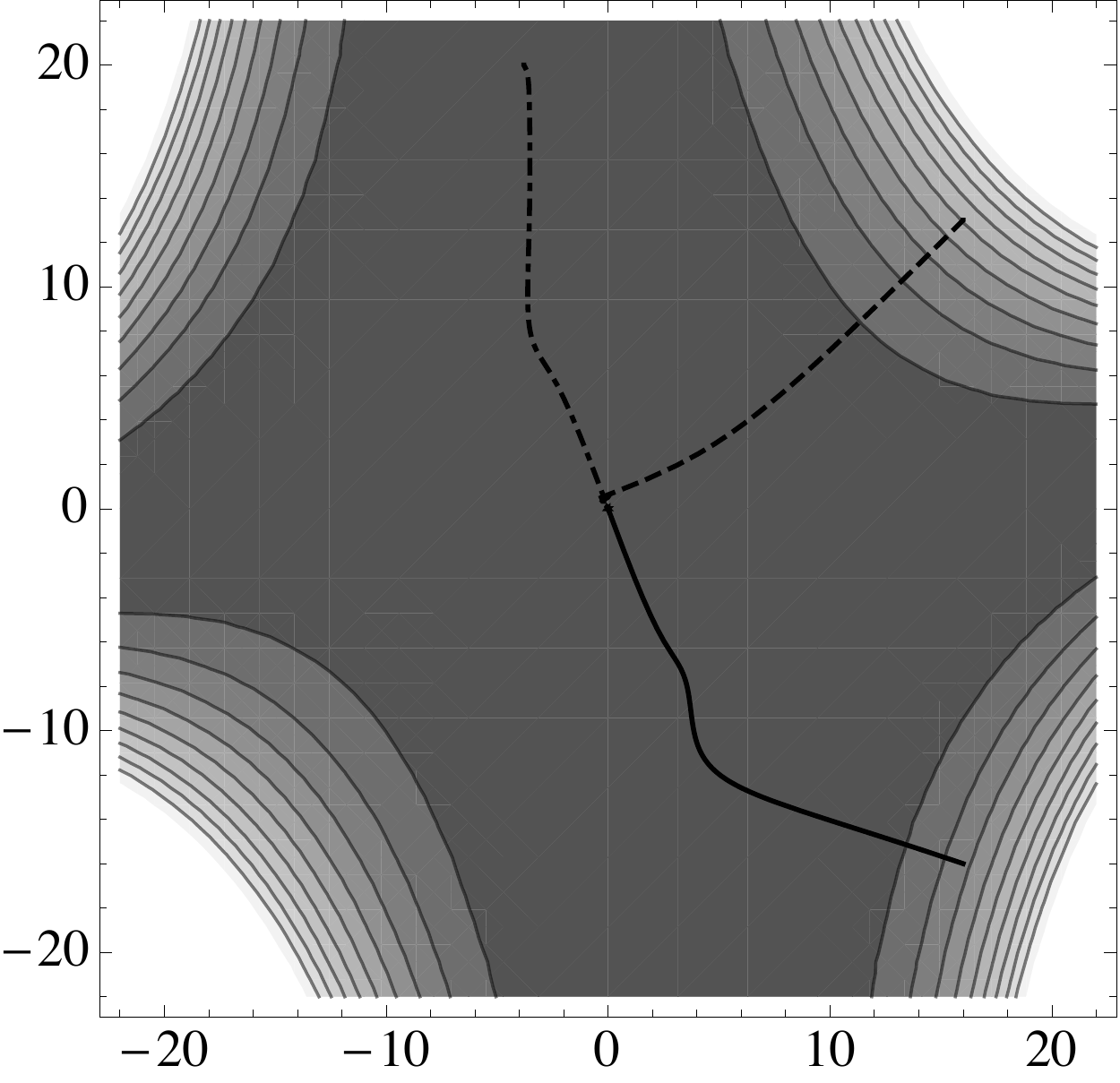} 
   \begin{picture}(0,0) 
   \put(-120,205){$\tilde \theta$}
   \put(-10,220){$\tilde \sigma$}
   \put(-195,260){$V$}
      \put(-82,-10){$\tilde \theta$}
   \put(-180,85){$\tilde \sigma$}
  \end{picture}
 % requires the graphicx package
   \caption{Plot of a) scalar potential and b) stable inflationary trajectories for $f = 1/40$, $h = 1/50$, $q = -1/17$ and $\lambda = 1/50$.}
   \label{fig:Vinf}
\end{figure}
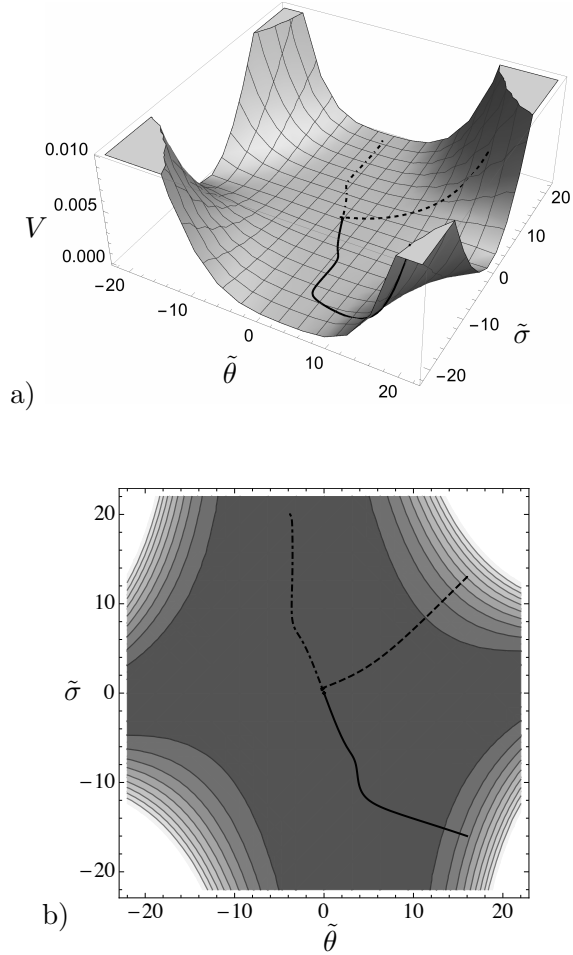

\section{Effective model of inflation}

We  proceed to compute the existence of three different inflationary trajectories\footnote{Their selection also follows from the number of e-folds they  produce as shown later on.}  shown in Figure \ref{fig:Vinf} in which  the scalar potential is presented as function of the axions $\tilde\theta$ and $\tilde\sigma$ for a specific choice of fluxes. Just for simplicity, we have selected $\Lambda=0$ although all our procedure can be  use to $\Lambda\ne 0$.
With the purpose of computing the number of e-folds that guarantees a required minimal amount of inflation, the initial points are selected such that they satisfy the inflationary conditions at the beginning of the trajectory while the pivot scales are chosen once the inflationary conditions are violated by reaching $\eta = 1$. The three initial values, the number of e-folds at the end of inflation and the selected fluxes are shown in Table \ref{tab:e-folds}.\\

% Requires the booktabs if the memoir class is not being used
\begin{table}[!htbp]
   \centering
   \begin{tabular}{@{} c | c @{}} % Column formatting, @{} suppresses leading/trailing space
      \hline
      $\left( \sigma_0, \theta_0 \right)$    & $e-folds$ \\
      \hline
      $\left( -3.8, 20 \right)$	  & 93 \\
      $\left( 16, 13 \right)$	  & 60 \\
      $\left( 16, -16 \right)$	  & 58 \\
      \hline
   \end{tabular}
   \caption{Initial value and number of e-folds obtained for selected trajectories.}
   \label{tab:e-folds}
\end{table}

The inflationary trajectories are obtained by a numerical solution of  Eqs. (\ref{scalarfieldmulti}) and (\ref{multiFriedman1}).  As usual we choose arbitrary points in the field space as the initial value and we solve numerically the FRW equations until the minima is reached. At the end of inflation ($\epsilon \sim 1$) we compute the number of e-folds. We propose three trajectories where at least 40--60 e--folds are obtained.  As we shall see, the expected hierarchy of scales is still preserved pointing out the possible presence of two-field inflation.

\subsection{Hierarchy of masses}

In this section we present the numerical values for the scales obtained in our model. The fluxes are selected in such a way that the saxions becomes heavier than its axion partners at least by one order of magnitude, and the two axions have a similar mass, this is
\eq{
m_{\rm s} > m_{\text{KK}} > m_\text{inf} \sim m_{\text{mod}} \sim  H_{\text{inf}} < m_{\tilde \theta} \sim m_{\tilde \sigma}   \,.
}
As is shown in Eq. \ref{eq:ms} mass of scales depend only on the saxions expectation values. Thus, for the present model the KK scale as well the string scale are fixed during inflation due to the hierarchy of saxions.  Table 2 shows the corresponding  values in Planck units.
%
% Requires the booktabs if the memoir class is not being used
\begin{table}[!htbp]
  \caption{Mass hierarchy for the selected model.}
   \label{tab:masshierarchy}
   \centering
   %\topcaption{Table captions are better up top} % requires the topcapt package
   \begin{tabular}{@{} cc @{}} % Column formatting, @{} suppresses leading/trailing space
      \hline % Partial rule. (r) trims the line a little bit on the right; (l) & (lr) also possible
      Mass    &  \\
      \hline
      $m_{\text{s}}$			& 5.64  \\
      $m_\text{KK}$         	&  $5.25 \cdot 10^{-2}$ \\
      $m_{\text{inf}}$		&  $1.257 \cdot 10^{-2}$\\
      $m_\text{saxions}$  & $\{ 9.99 \cdot 10^{-3} \,, 2.23 \cdot 10^{-2} \}$ \\
      $H_{\text{inf}}$		& $2.00 \cdot 10^{-2}$ \\
      $m_\text{axions}$    & $\{ 1.75 \cdot 10^{-3} \,, 2.37 \cdot 10^{-4} \}$\\
      \hline
   \end{tabular}
\end{table}

Tadpole cancellation condition for the selected models implies
\eq{
\mff h = N_{O3}-N_{D3} \,, \quad  \mff q = N_{O7}-N_{D7} \,,
}
which for the selected fluxes requires $N_{O3}>N_{D3}$ and $N_{O7}<N_{D7}$.  Notice that in our numerical example all fluxes have values less than 1. Therefore their contribution to tadpoles is of order $\mathcal{O} \left( 10^{-4} \right)$ in Planck units. Although fractional amounts of fluxes are not allowed by Dirac quantization, the fluxes in the superpotential could have perturbative and non-perturbative corrections as shown in \cite{Hosono:1994av}.  In such case, no tadpole exists in the model. Namely, in the limit of large complex structure, the prepotential is modified by perturbative and non-perturbatibe corrections coming from the mirror as
\eq{
\mathcal{F} = \frac{1}{6}\kappa_{ijk}\frac{z^i z^j z^k}{z_0}+ \frac{1}{2}a_{ij}z^i z^j + b_i z^i z^0 + \frac{1}{2}c(z^0)^2 +\mathcal{F}_{\text{inst}}.
}
where $\kappa$ is the triple intersection number, $a_{ij}$ and $b_i$ are perturbative corrections to the superpotential whereas $\gamma \in \mathbb{R}$ which corresponds to $\alpha$' corrections not considered in the present model shall be safely neglected as well to the non-perturbative corrections $\mathcal{F}_{\text{inst}}$. These, contributions generates a rational contribution to the superpotential that mimics the behavior of the fluxes.\\

As pointed out by \cite{Ooguri:2016pdq} all the vacua coming from flux compactification in which there exists branes charged under fluxes as those considered here,  are unstable unless they are protected by a BPS condition. However, in the present example no such branes are expected to appear. This is, in order to get inflation it is required to select fractional fluxes that can be understood as non-perturbative corrections to the complex structure of the mirror. In the case when no fluxes are turned on, the non-perturbative corrections modifies the superpotential and in an effective manner its contributions mimics the behavior of the fluxes stabilizing the moduli. This interesting question shall be postponed for future work.

 \subsection{Spectral index}
\begin{figure}[!htbp]
   \centering
   a) \includegraphics[width=7cm]{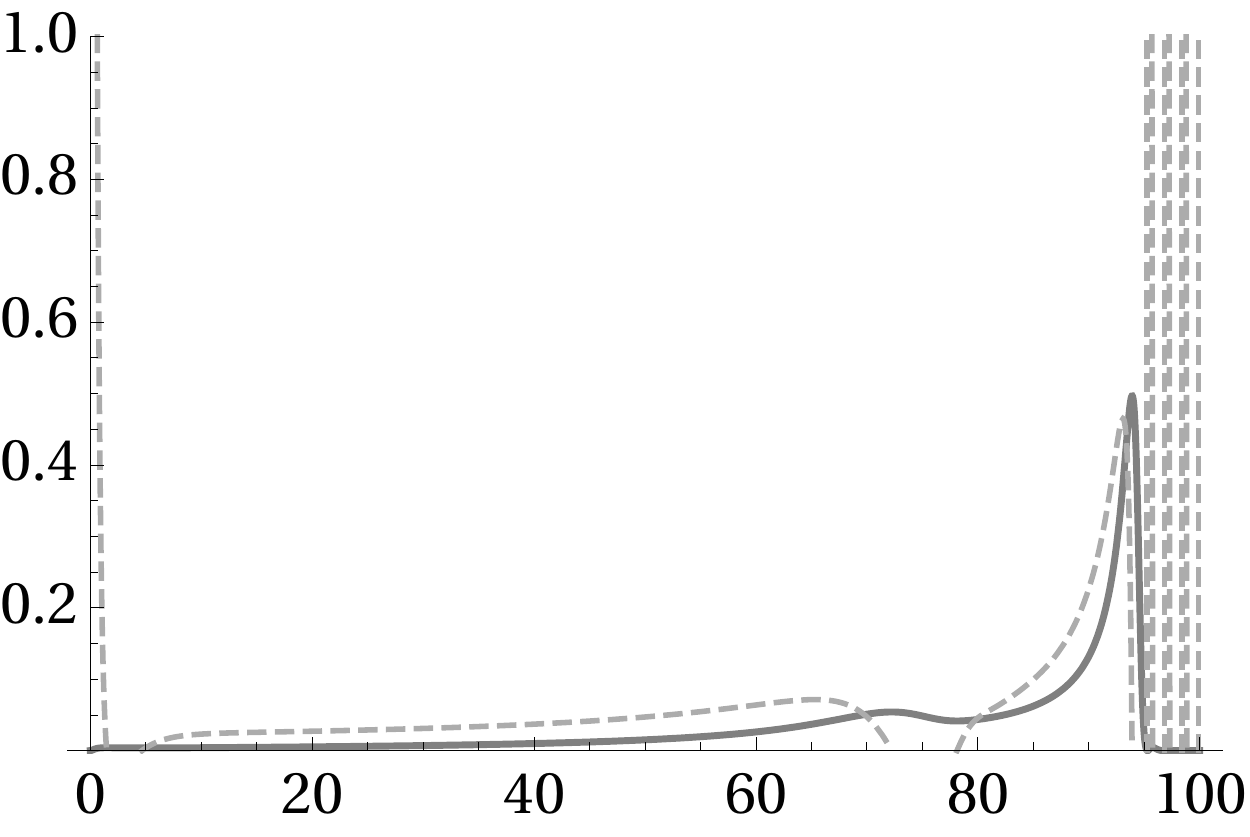} \\ b) \includegraphics[width=7cm]{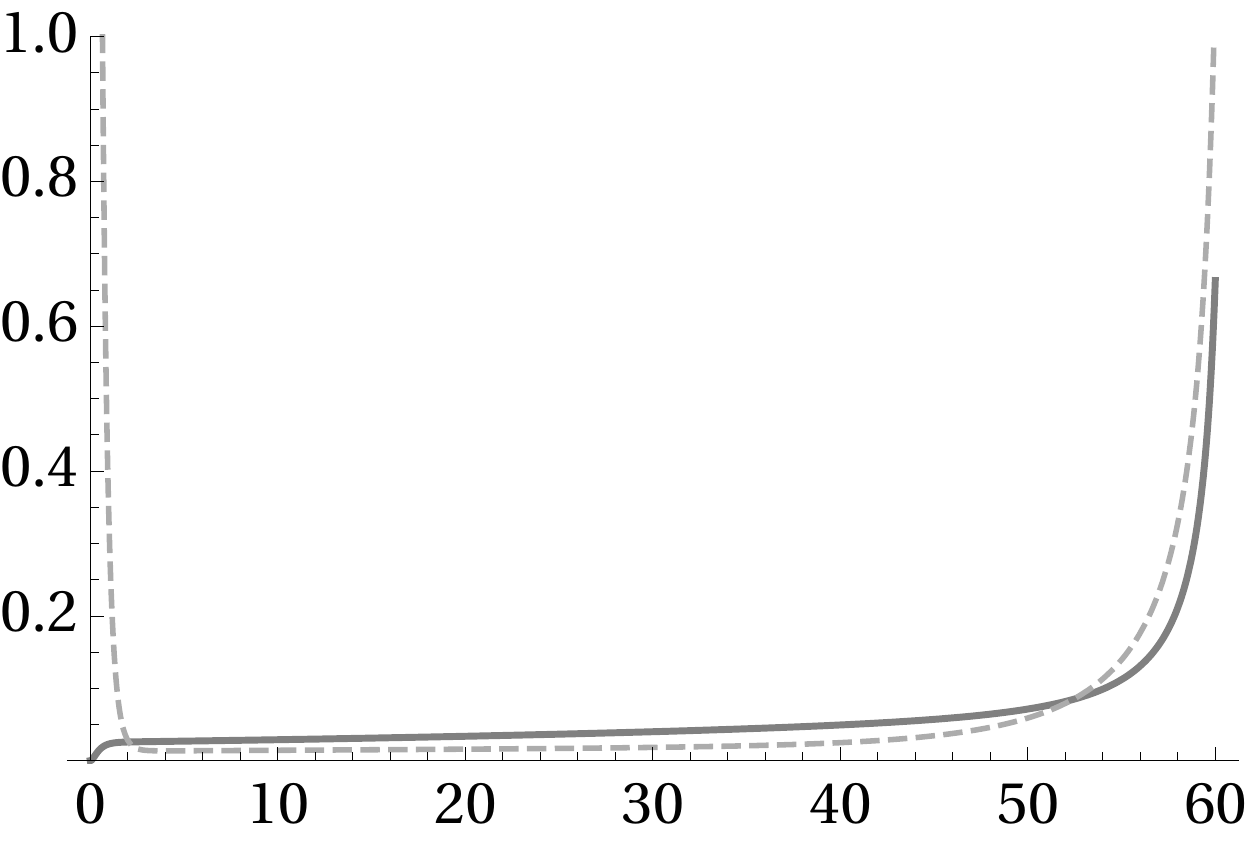}\\
   c) \includegraphics[width=7cm]{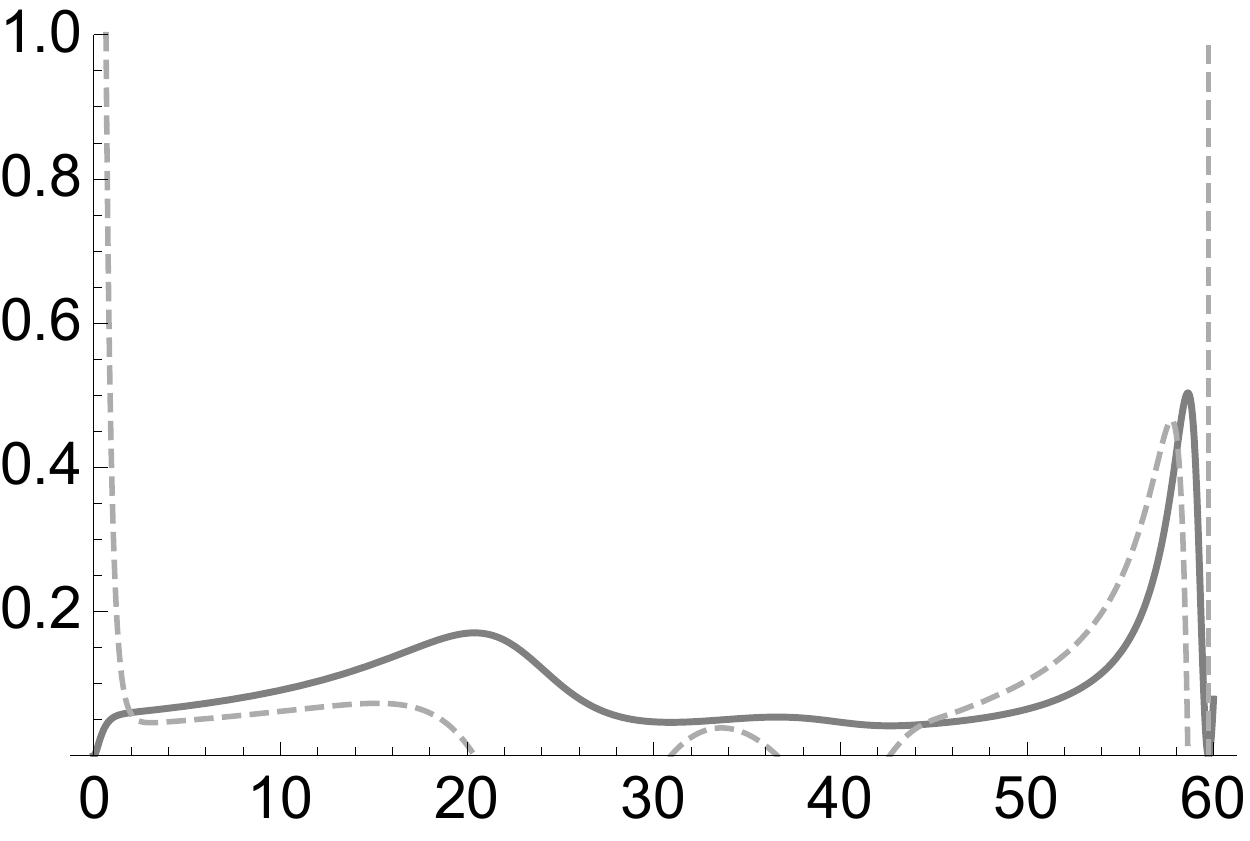} % requires the graphicx package
   \begin{picture}(0,0)
   \put(0,10){$\rm N_e$}
   \put(-180,125){$\epsilon$, $\eta$}
   \put(0,150){$\rm N_e$}
   \put(-180,260){$\epsilon$, $\eta$}
   \put(0,290){$\rm N_e$}
   \put(-180,400){$\epsilon$, $\eta$}
   
    \end{picture}
   \caption{Slow--roll parameters $\epsilon$(solid line) and $\eta$ (dashed line) for  a) trajectory I, b) trajectory II and c) trajectory III.}
   \label{fig:eeta}
\end{figure}
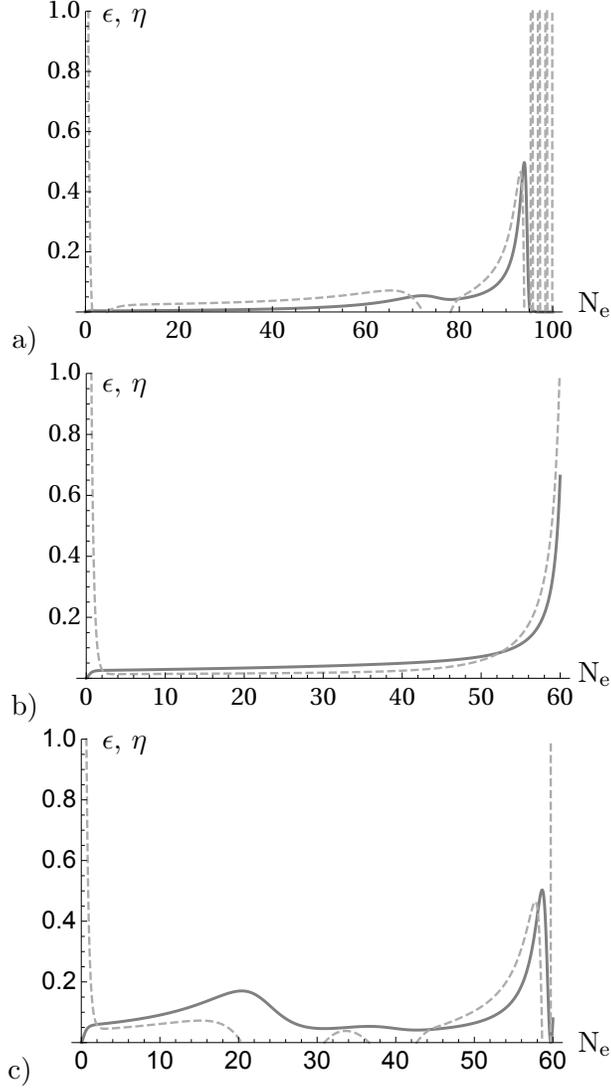

 In Figure \ref{fig:eeta} it is shown the evolution of the slow-roll parameters as a function of the e-folding.  As observed for all the trajectories, at the beginning of inflation there exists a momentarily fast-rolling region which does not last more than a couple of e-folds.  Besides that we find that there exist a range in moduli space in which inflation occurs. A particularly interesting issue appears for trajectory III
which contains a region in the field space where there is a slightly accelerated expansion while fulfilling the slow-roll conditions. As we shall see, this region has important contributions to all  cosmological observables.\\

\begin{figure}[htbp]
   \centering
   a) \includegraphics[width=7cm]{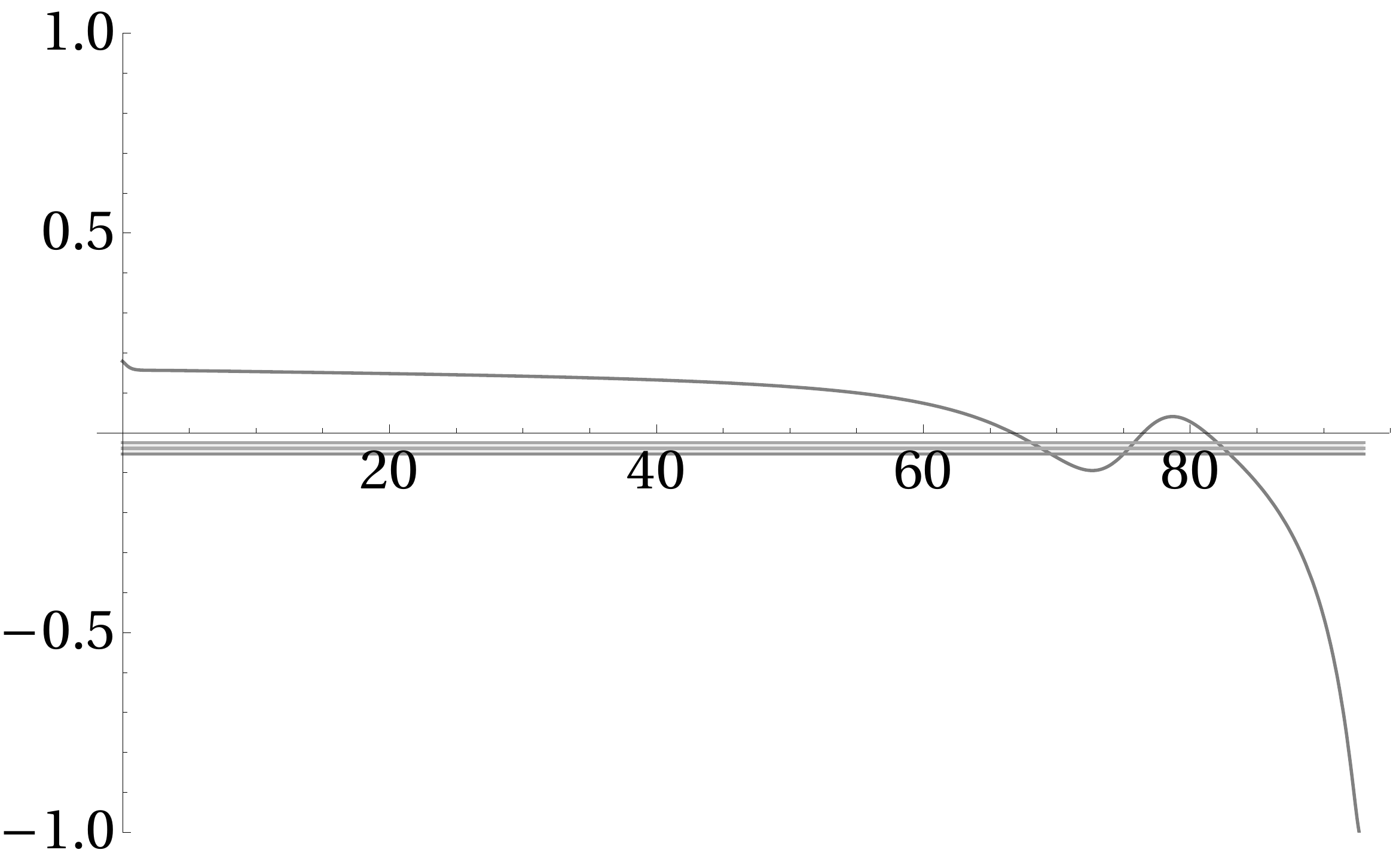} \\ b) \includegraphics[width=7cm]{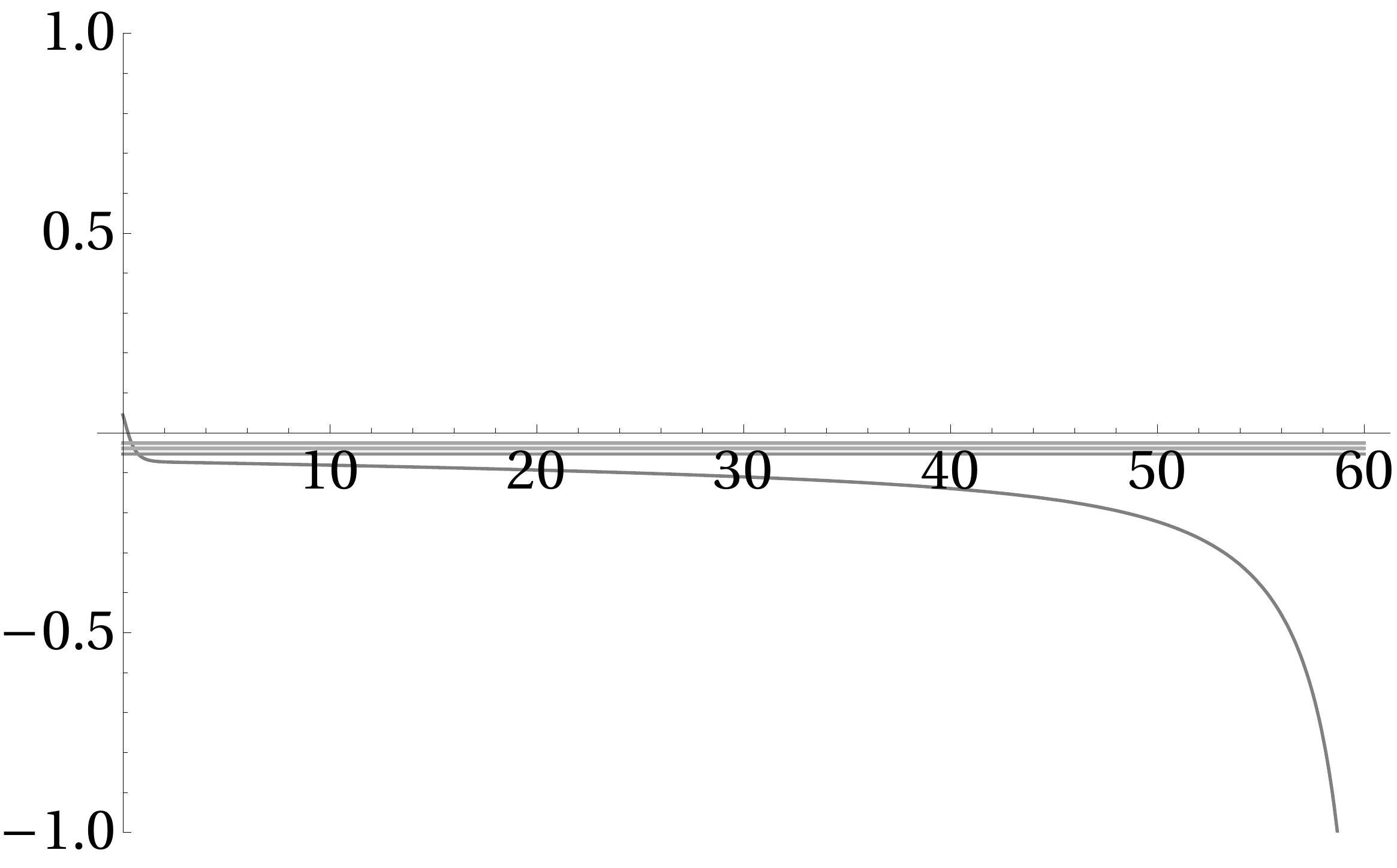}\\
   c) \includegraphics[width=7cm]{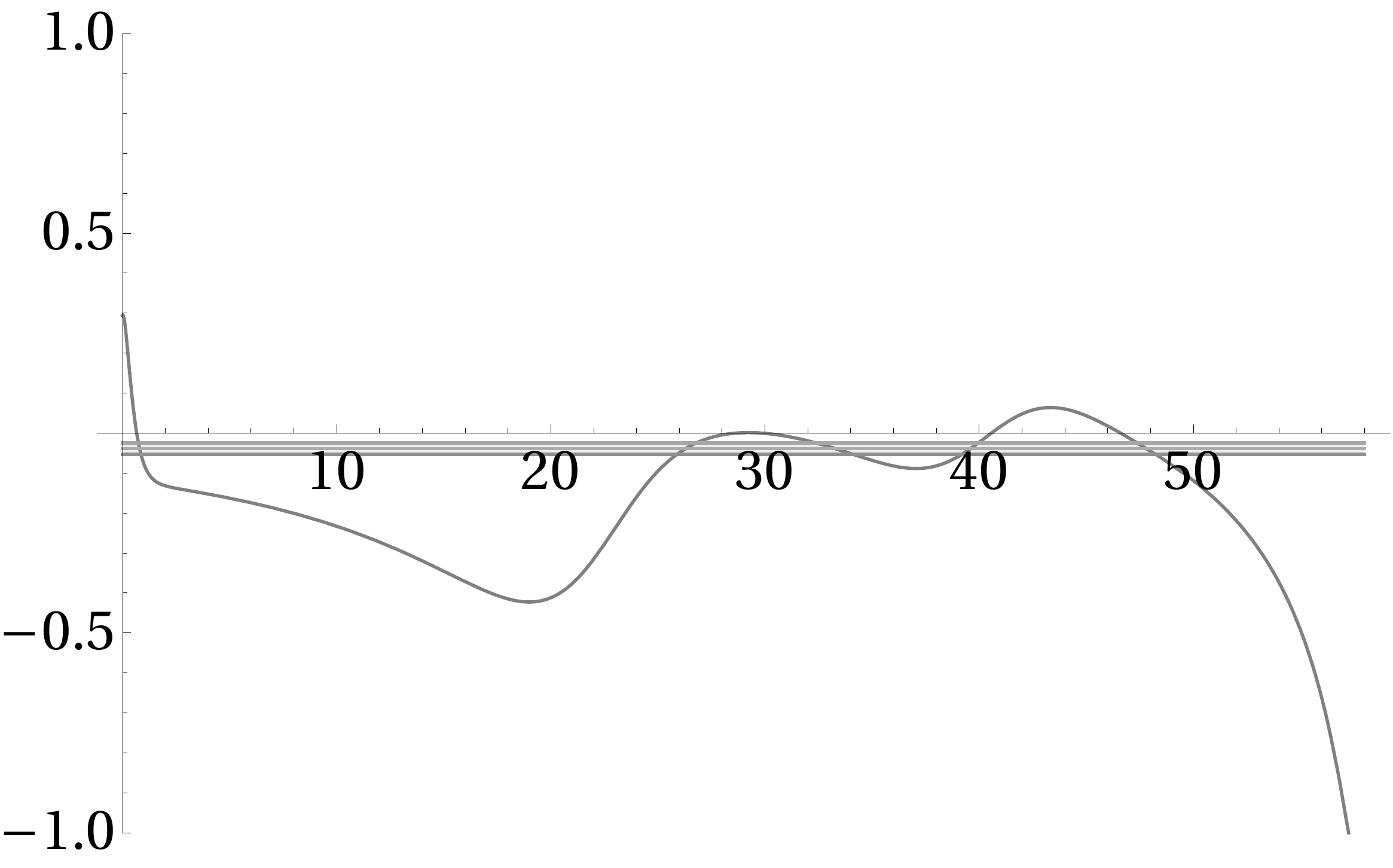} % requires the graphicx package
    \begin{picture}(0,0)
   \put(0,63){$\rm N_e$}
   \put(-178,120){$n_s-1$}
    \put(0,190){$\rm N_e$}
   \put(-178,245){$n_s-1$}
   \put(0,315){$\rm N_e$}
   \put(-178,370){$n_s-1$}
    
  \end{picture}
   \caption{Plot of spectral index $n_s$ for a) trajectory I, b) trajectory II and c) trajectory III.}
   \label{fig:nsr}
\end{figure}

 The spectral index $n_s$ is calculated from the relation shown in Eq. \ref{ns} and it is shown in Figure \ref{fig:nsr} for our three trajectories. The filled lines represent the allowed value of $n_s$ as reported by Planck.
 %. 
The results shown that all trajectories meet the observable bounds at different times during the slow-roll dynamics. For the case of trajectory I, the  bounds are matched at the end of inflation.  For trajectories I and III the horizon exit is expected to occur at the end of inflation in order to agree with observations. However, for trajectory II, observational bounds are matched at the beginning of inflation. As is pointed out in the previous paragraph, there is a region in the field space where the spectral index varies in a oscillatory manner. This region is related to the variation of the $\epsilon$ parameter. This behavior implies an acceleration during the slow-roll regime. 

\subsection{Tensor to scalar ratio}
\begin{figure}[htbp]
   \centering
   a) \includegraphics[width=7cm]{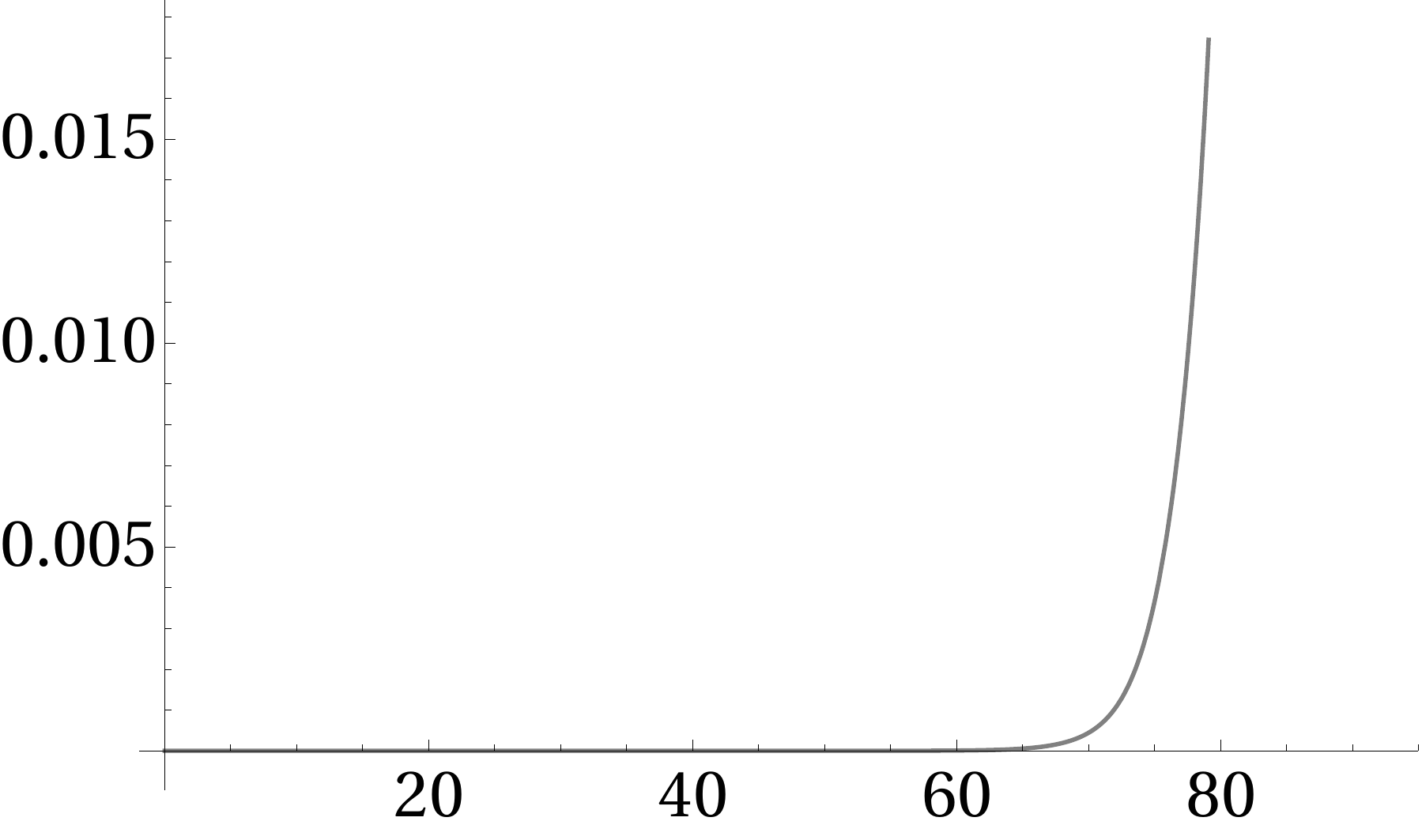}\\ b) \includegraphics[width=7cm]{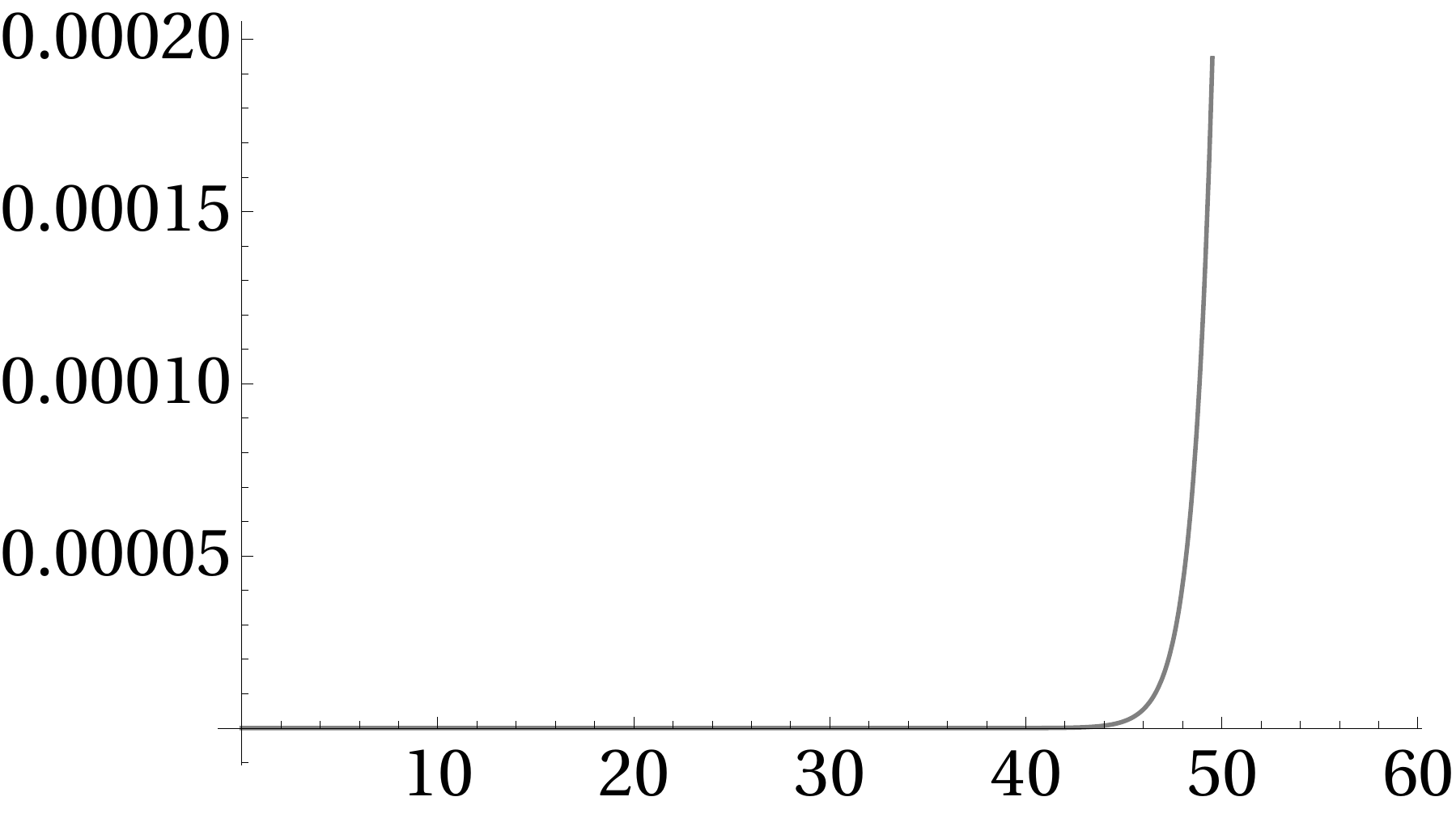}\\
   c) \includegraphics[width=7cm]{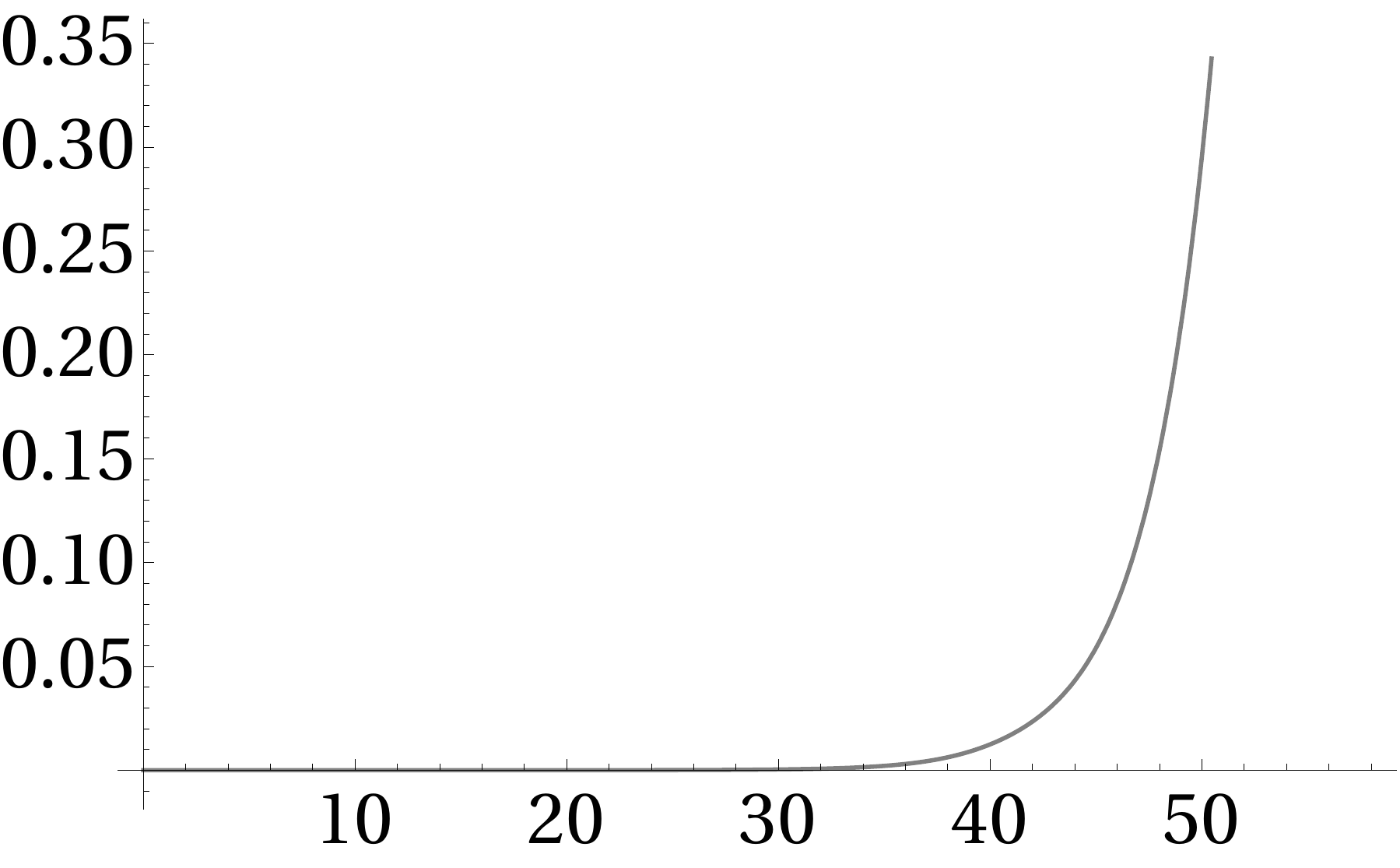} % requires the graphicx package
   \begin{picture}(0,0)
   \put(-5,15){$\rm N_e$}
   \put(-177,117){$r$}
   \put(-5,138){$\rm N_e$}
   \put(-165,235){$r$}
   \put(-5,255){$\rm N_e$}
   \put(-170,355){$r$}

  \end{picture}
  \caption{Plot of the tensor to scalar ratio $r$ for a) trajectory I, b) trajectory II and c) trajectory III.}
   \label{fig:r}
\end{figure}

The tensor to scalar ratio $r$ is calculated through the Eq. \ref{r} which is defined as the ratio of power spectra of tensor and scalar perturbations. Since the claim of the possibility to have scenarios with a non-negligible value of $r$, this cosmological observable has attracted considerable attention in the last years.  In Figure \ref{fig:r} we present the tensor to scalar ratio for the three selected trajectories. As observed the trajectory III presents a large ratio at the end of inflation which does not meet the current observational bounds of $r<0.05$ as reported by Planck. However, trajectory I and II, give a small value of $r$ which are in good agreement with the reported values.

\subsection{non-Gaussianities}
The local non-Gaussianities $f_{NL}$ are determined through Eq. \ref{fnl} as a position space expansion around Gaussian perturbations. Under this ansatz, non-Gaussianities are generated independently at different spatial points \cite{Rubin:2001yw,Khlopov:2002yi,Wang:2013eqj}. Translating this requirement into the context of inflation, typically  implies that non-Gaussianities are generated on super-Hubble scales. Therefore,
\begin{equation}
\xi (x) = \xi_g (x) + \frac{3}{5} f_{NL} \xi^2 (x) + \ldots
\end{equation}
where $\xi_g (x)$ is a perturbation variable that satisfies the Gaussian statistics. Thus a positive value of local non-Gaussianities enhance the value of the power spectra, providing more hot spots on the CMB, whereas an negative value shall be related to colder spots. Thus, a positive value of $f_{NL}$ shall be related to more large structure formation of Galaxies.
 \begin{figure}[htbp]
   \centering
   a) \includegraphics[width=7cm]{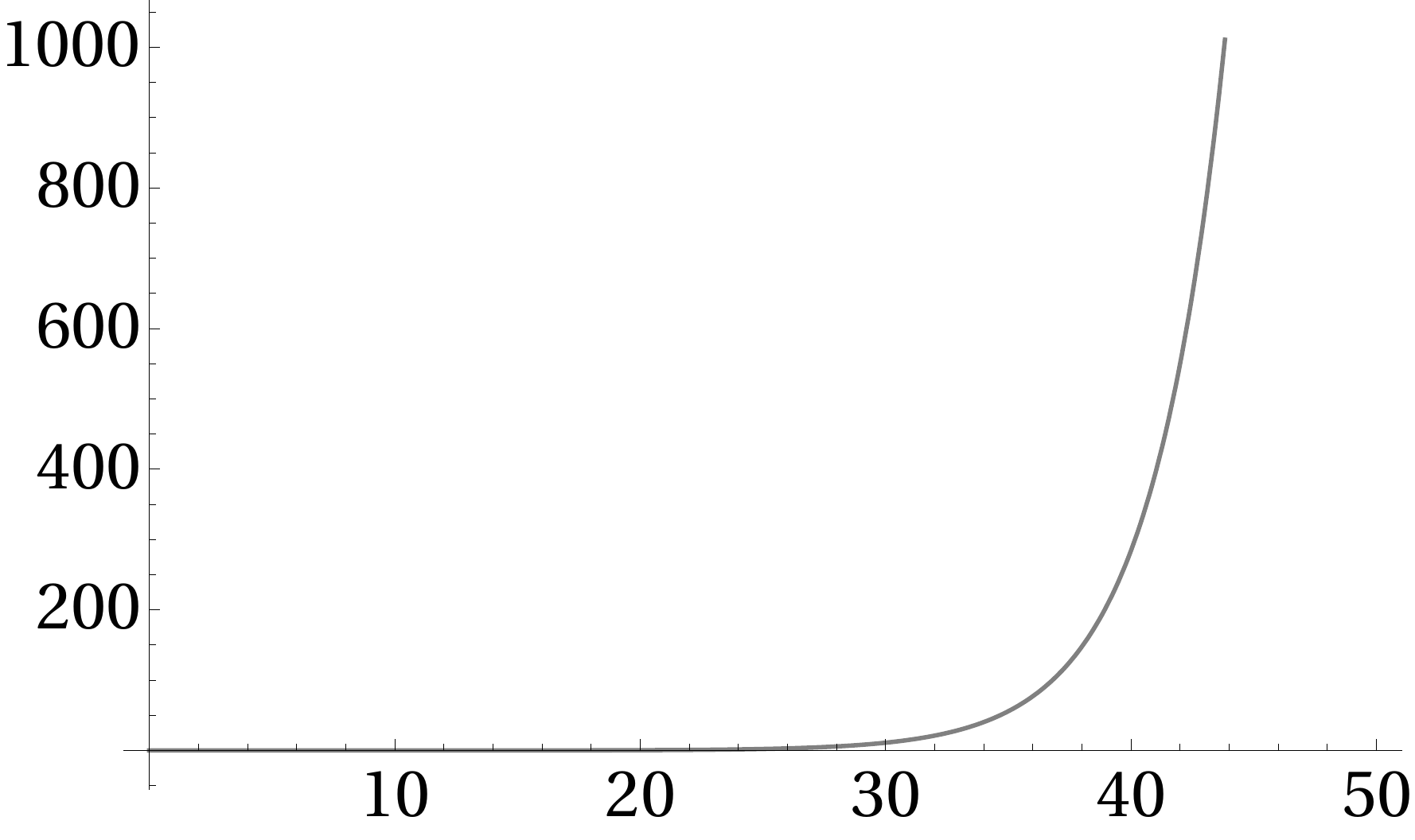} \\ b) \includegraphics[width=7cm]{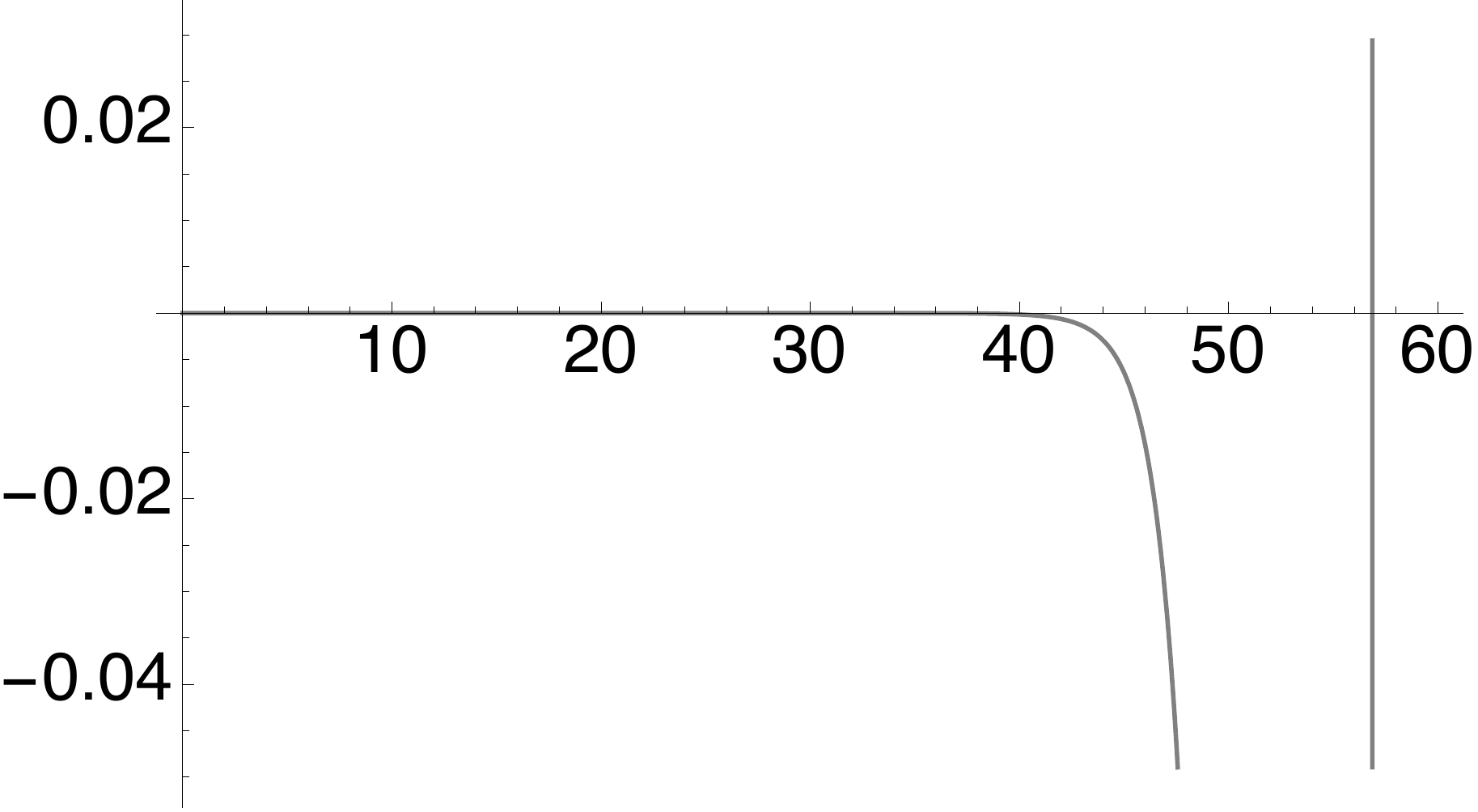}\\
   c) \includegraphics[width=7cm]{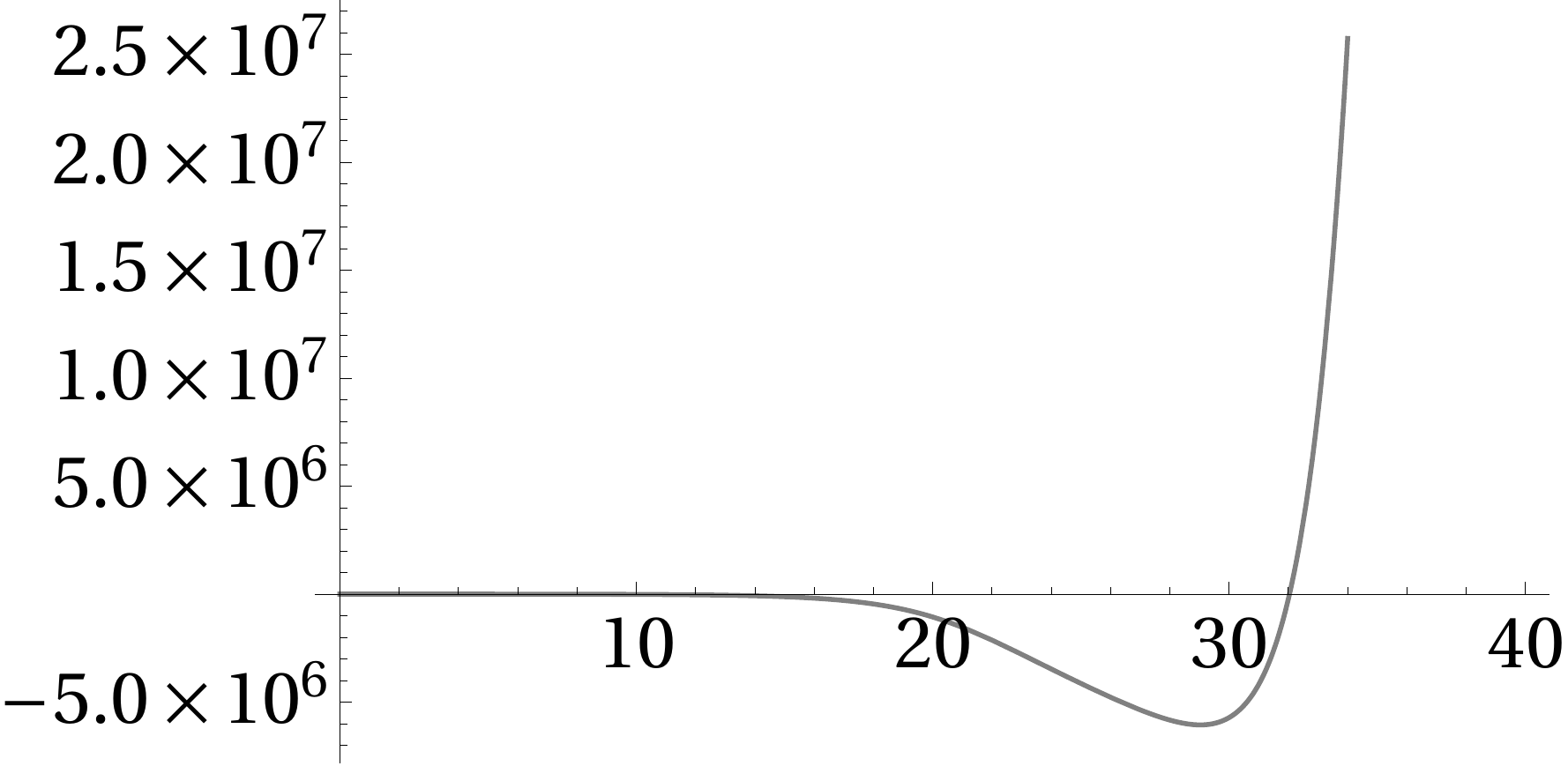} % requires the graphicx package
    \begin{picture}(0,0)
   \put(0,20){$\rm N_e$}
   \put(-155,90){$f_{NL}$}
   \put(0,170){$\rm N_e$}
   \put(-173,200){$f_{NL}$}
   \put(0,225){$\rm N_e$}
   \put(-175,325){$f_{NL}$}
  
  \end{picture}
  \caption{Plot of non-Gaussianities for a) trajectory I, b) trajectory II and c) trajectory III.}
   \label{fig:fNL}
\end{figure}
Figure \ref{fig:fNL} shows  non-Gaussianities obtained in the presented model for our three trajectories. As observed,  along trajectory II we find $|f_{NL}| < 0.1$ giving rise to a nearly Gaussian spectrum, which is compatible with the current observation. However, for trajectories I and III, the model predicts a very large amount of non-Gaussianities and in consequence a large amount of hot(blues) spots in the CMB which are beyond the current observational bounds. \\

\section{The swampland criterion}
Construction of realistic scenarios from string compactifications has faced a variety of obstacles and efforts to overcome them. It is difficult to  distinguish effective field theories obtained from string theory by some type of compactification to those which apparently are not compatible with a quantum theory of gravity. The landscape and the swampland, respectively, are two regions in which stringy models may belong to. Recently, a criterion for that was presented in \cite{Obied:2018sgi, Agrawal:2018own} establishing a bound from below to the gradient of the scalar potential in case it has a positive value.  The proposal has received a lot of attention specially for its implications on cosmology  \cite{Andriot:2018wzk, Dvali:2018fqu, Garg:2018reu, Kehagias:2018uem, Denef:2018etk, Roupec:2018mbn, Andriot:2018ept, Matsui:2018bsy, Ben-Dayan:2018mhe, Dias:2018ngv}  which essentially states that single-field slow-roll inflation is ruled out.  However, for the case of a multi-field inflationary scenario it was shown in  \cite{Achucarro:2018vey}  that the above criterion and inflation can in fact coexist. Therefore, we want to study wether our model is compatible with the swampland criterion while preserving inflationary trajectories. For that we start with a very brief review on the swampland proposal in a multi-field scenario closely following \cite{Achucarro:2018vey}.

\subsection{Multi-field inflation and the swampland constraint}
The inflationary swampland criterion about the bounding on the slope of the scalar potential is given by
\eq{
\frac{| \nabla V|}{V}\geq c\sim {\cal O}(1),
}
which for a single field effective theory implies that the slow-roll parameter $\epsilon\sim {\cal O}(1)$ ruling out inflation. However, as shown in \cite{Achucarro:2018vey}, the above criterion must be generalized for the case of a multi-field scenario in which the bound is now given by  $\epsilon_V$, with
\eq{\label{vafa}
\epsilon_V=\epsilon\left(1+\frac{\Omega^2}{9H^2}\right).
}
The parameter  $\Omega$ measures the bending of the trajectory with respect to a geodesic in the moduli field space. Essentially, is given by the modulus of the covariant derivate on time of the tangent component of the trajectory. Therefore, if the ratio $\Omega/3H$ is large enough to produce an equally large value of $\epsilon_V$ the swampland criterion is fulfilled and inflation is also present on such trajectory if $\epsilon$ is still small. It was then argued in \cite{Achucarro:2018vey} that for non-geodesic trajectories it is possible to have inflation with a large $\epsilon_V$. In this section we shall show that this is indeed the case for our model.

\subsection{The swampland criterion for two-field inflationary model in the flux scaling scenario}

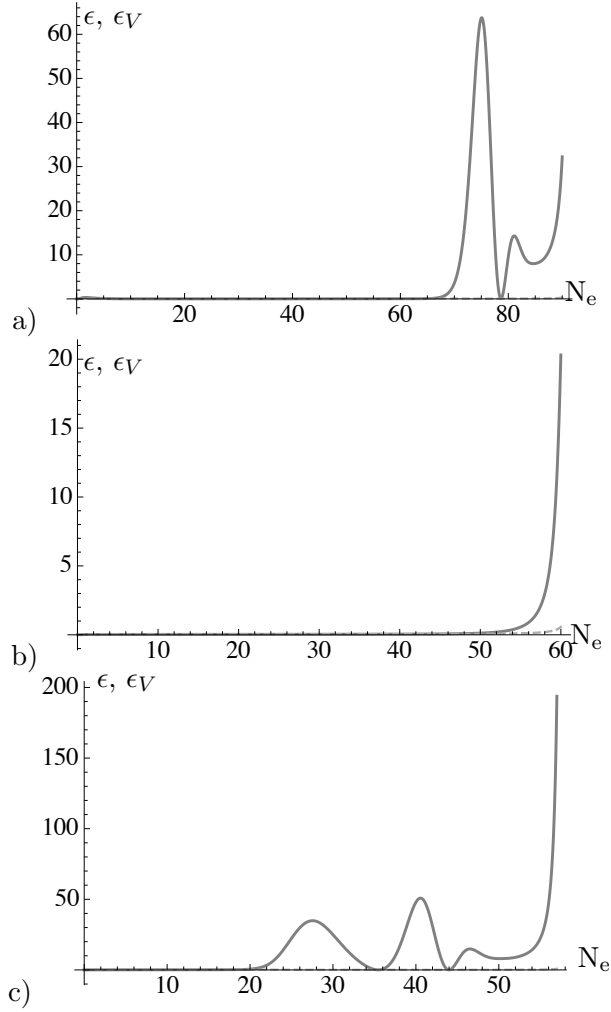
\begin{figure}[htbp]
   \centering
   a) \includegraphics[width=7cm]{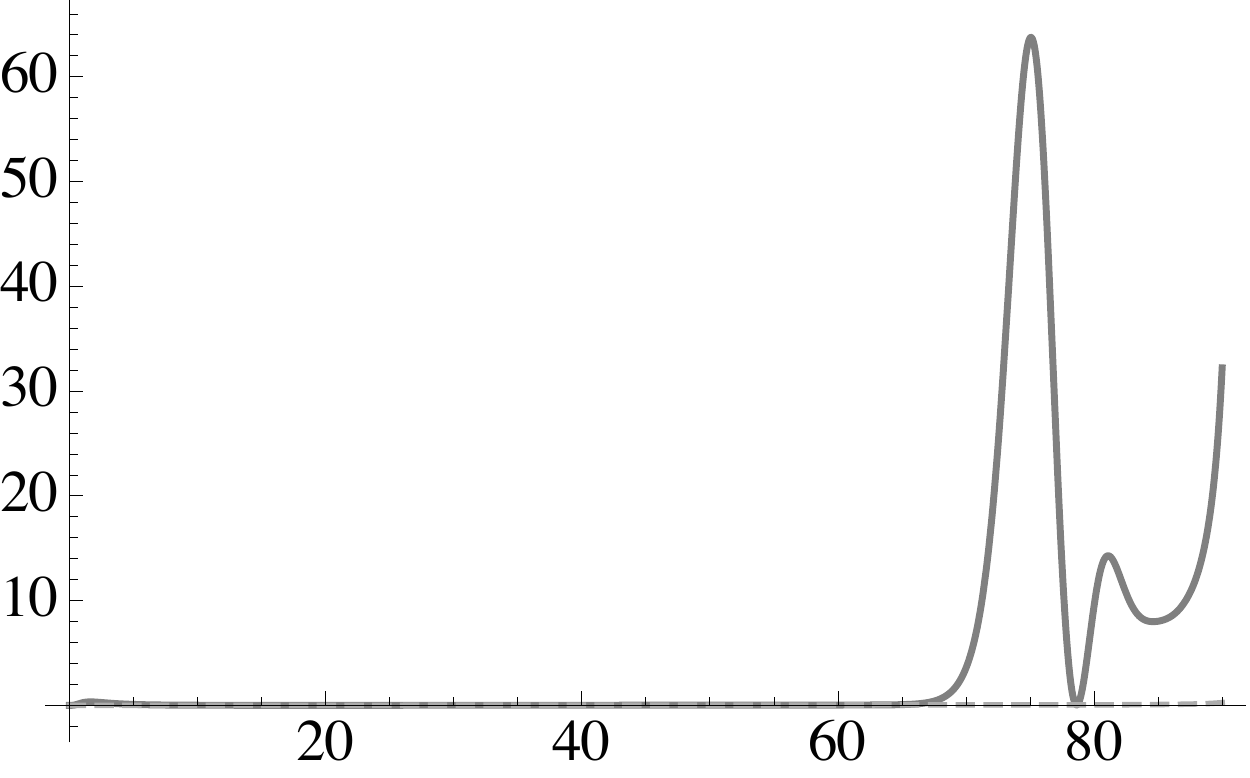} \\
   b) \includegraphics[width=7cm]{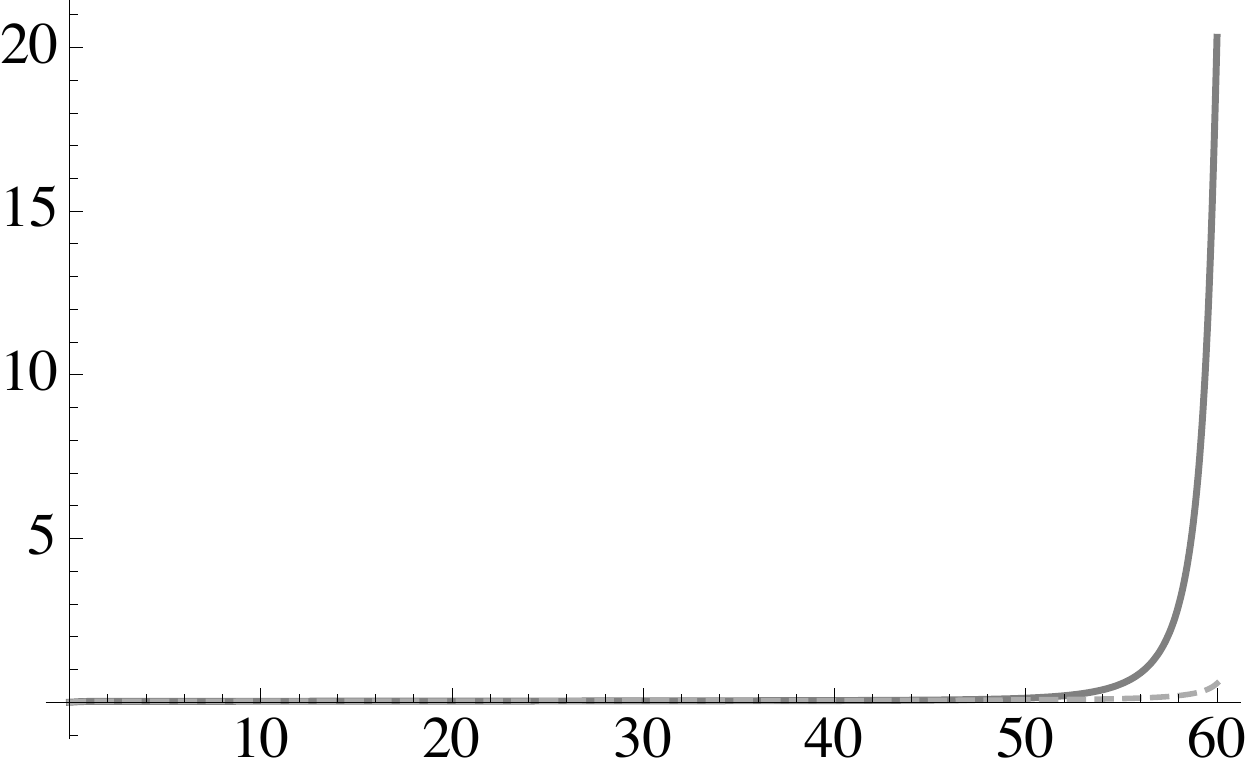}\\
   c) \includegraphics[width=7cm]{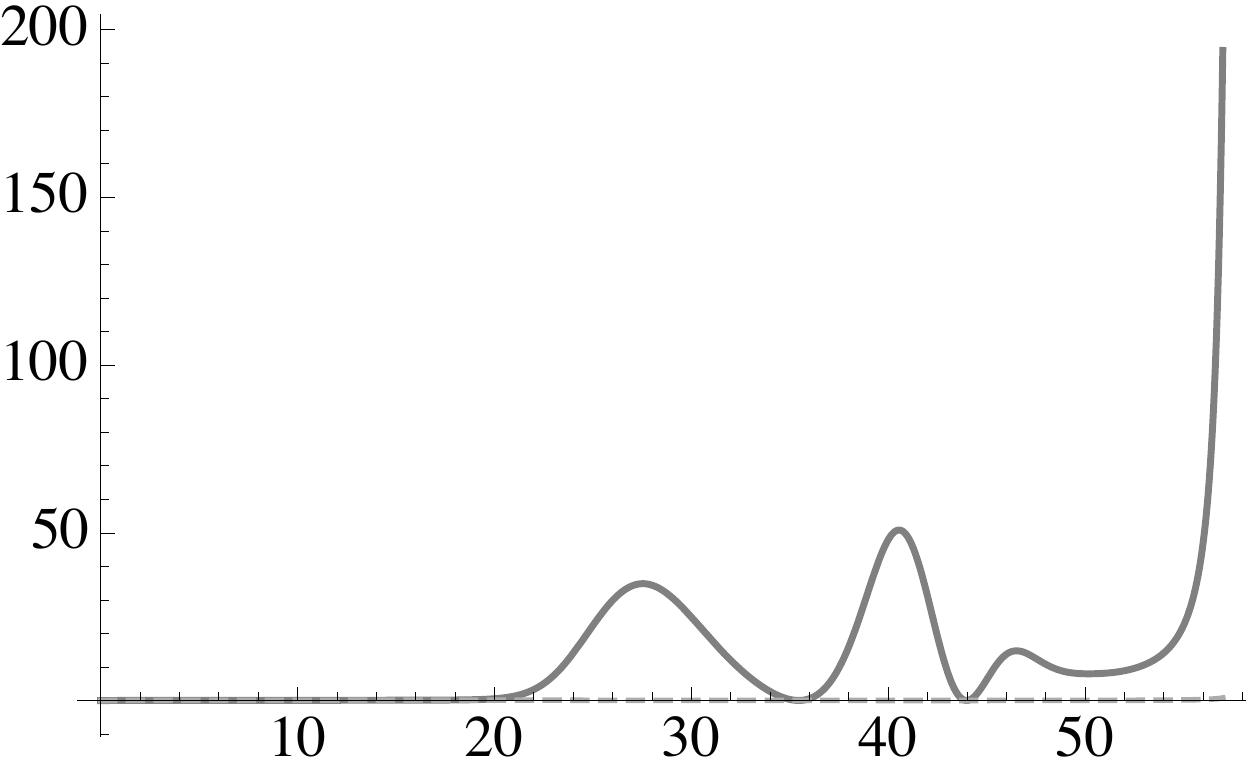} % requires the graphicx package
    \begin{picture}(0,0)
   \put(0,13){$\rm N_e$}
   \put(-182,119){$\epsilon$, $\epsilon_V$}
    \put(-3,135){$\rm N_e$}
   \put(-187,239){$\epsilon$, $\epsilon_V$}
   \put(-5,265){$\rm N_e$}
   \put(-187,370){$\epsilon$, $\epsilon_V$}

  \end{picture}
  \caption{Plot of inflationary parameters $\epsilon$ and $\epsilon_V$ for a) trajectory I, b) trajectory II and c) trajectory III.}
   \label{fig:swampland}
\end{figure}

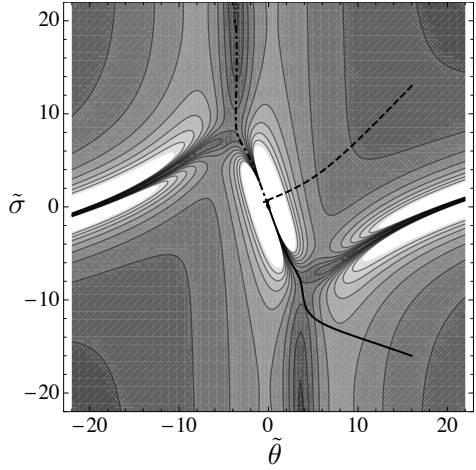
\begin{figure}[htbp]
   \centering
    \includegraphics[width=6cm]{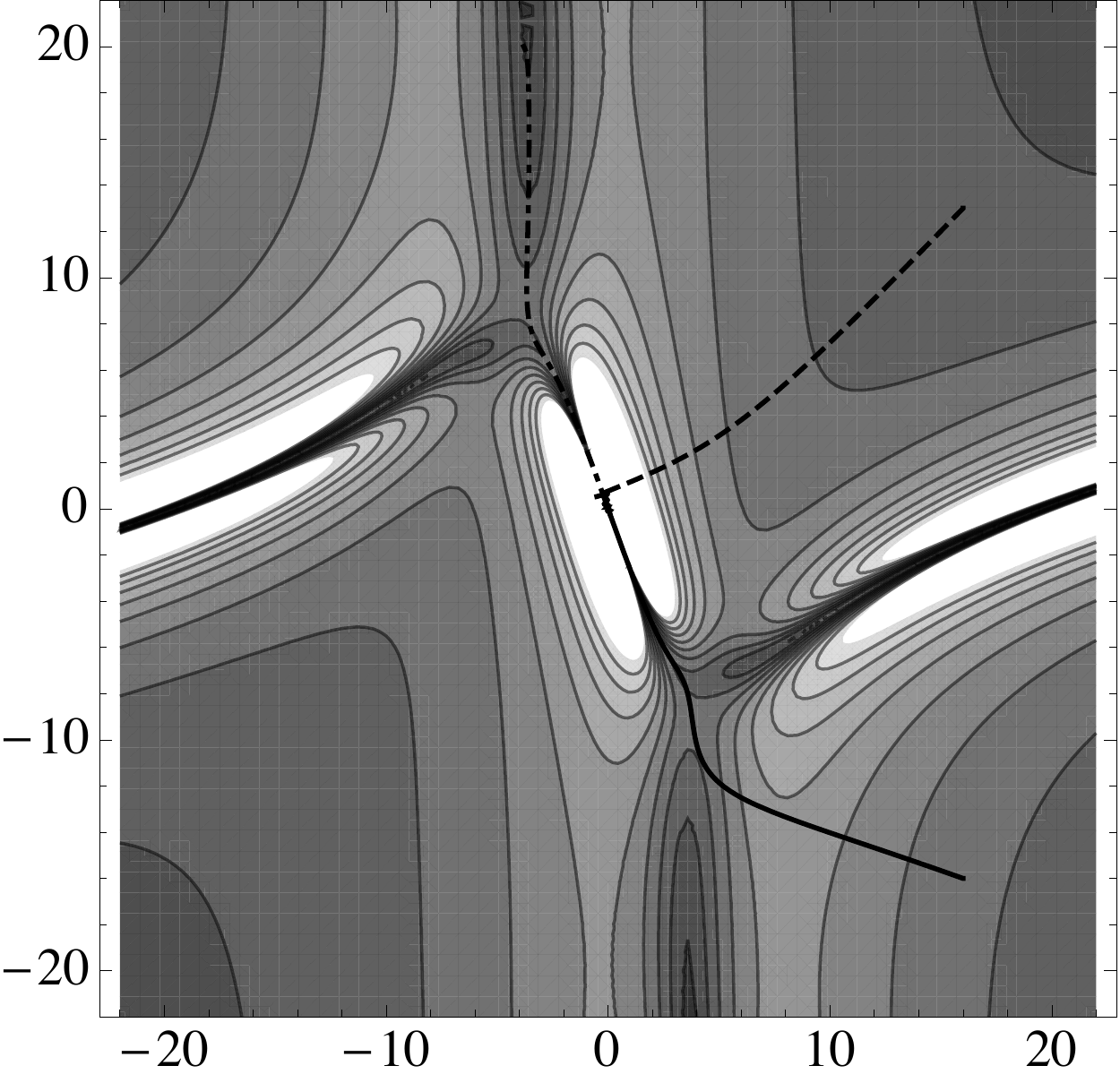}
   \begin{picture}(0,0) 
   \put(-82,-10){$\tilde \theta$}
   \put(-180,85){$\tilde \sigma$}
  \end{picture}
 % requires the graphicx package
   \caption{Contours of $|\nabla V|/V$ keeping the saxions at the minima for $f = 1/40$, $h = 1/50$, $q = -1/17$ and $\lambda = 1/50$.}
   \label{fig:boundvafa}
\end{figure}

In Figure \ref{fig:swampland} we compare $\epsilon$ and $\epsilon_V$ for the three selected models studied in the previous section. As is shown in \cite{Achucarro:2018vey}, there is a deviation from $\epsilon$ due to the fact that we have a two-field inflation scenario. This is, whereas $\epsilon$ remains small during all the trajectory, there exists variations for $\epsilon_V$ due to angular velocity on the field trajectory  which makes it bigger than 1.  It is however, important to mention that these variations do not fulfill the bound proposed in \cite{Achucarro:2018vey} about having $\epsilon_V> 180H$.   For trajectories 1 and 3  we see that variations occur along the trajectory and not just at the end as it happens to trajectory 2. It seems to be some relationship with the contours given by the bound $|\nabla V|\geq cV$ and such variations.
For sake of completeness let us try to construct an explicit form for $\Omega$.  For that we consider the superpotential related to a Minkowski vacuum, this is, we shall take $\Lambda=0$. Some comments about $\Lambda\ne 0$ are given  at the end of this section. Once we have the stabilized values for $\tilde\theta$ and $\tilde\sigma$, the scalar potential can be written as a function of $s$ and $\tau$ and also of the fluxes $h, f$ and $q$. Now we take the contribution to the scalar potential by turning on the flux $p$ by which  we obtain the vev's for $s$ and $\tau$ given by 
\eq{
s &=\frac{1}{2^3} \frac{f}{h} + \frac{47}{2^5} \frac{p f}{q h} c + \frac{2}{3} \frac{p \theta}{q f} c^2 - \frac{7}{2^7} \frac{p f \theta}{q h} \,,  \\
\tau &= -\frac{3^2}{2^3} \frac{f}{q} - \frac{303}{2^5} \frac{p f}{q h} c + \frac{59}{2^7} \frac{p f \theta}{q^2 h} - \frac{2\cdot 7}{3} \frac{p h \theta}{q^2 f} c^2 \,.\\
}
For small values of $p$ and $\tilde\theta$ the scalar potential is written at first order on $p$ as
\eq{
V_{eff} = \frac{2^2}{\sqrt{7}}\frac{p q^2}{f} \left( \frac{13}{3^3} \tilde \theta - \frac{1}{3^5}\tilde \theta^3 - \frac{2}{\sqrt{3}}\tilde \sigma + \frac{2}{3^3\cdot\sqrt{3}}\tilde \sigma \tilde{\theta}^2 -\frac{1}{3^2}\tilde \sigma^2 \tilde \theta \right)   \,.
}
This effective potential reflects some interesting properties: first of all inflationary directions are driven by $\tilde\theta$ and $\tilde\sigma$ while the rest of axions remain fixed at their vevs. This is accomplished by construction where slow-roll inflation is present due to the non-integers fluxes and a non-constant $p$-flux. Second of all, we see that $\epsilon_V$ grows to values bigger than 1 along inflationary trajectories where $\epsilon$ remains smaller than unity. The swampland bound given in  (\ref{vafa}) for this scalar potential $V_{eff}$ is fulfilled in zones very close to the dS minimum as shown in Figure (\ref{fig:boundvafa}). We also observe that those zones in $\epsilon_V$ is bigger than one are not the same as those in which $|\nabla V|\geq c V$ since the former is also measuring the angular momentum along the trajectory. However as mentioned before, $\epsilon_V$ is still below the minimal bound suggested in \cite{Achucarro:2018vey} of 180 H.

In order to completely geometrize the swampland criteria, let us determine the dependence of the $\Omega$ parameter in the slow-roll approximation in terms of the involved fluxes. This will allows us to see wether it is possible to have some kind of parametrical or numerical control on $\Omega$ by the fluxes. First, we notice that $\Omega$ is related to the dynamics of the scalar fields as
\eq{
\Omega = \frac{\dot{\tilde \sigma}\ddot{\tilde \theta}-\dot{\tilde \theta}\ddot{\tilde \sigma}}{\dot{\tilde \theta}^2 +\dot{\tilde \sigma}^2} 
}
and the time derivative of the Hubble scale is determined by
\eq{
\dot H = -\frac{1}{2} \dot \phi_0^2 \,,
}
for $\phi_0 = \sqrt{\delta_{ij} \dot{\phi}^i \dot{\phi}^j}$, thus we have that the time variation of the Hubble scale is related to the kinetic energy of the scalar fields as
\eq{
\dot H = -\frac{1}{2} \left( \dot{\tilde \theta}^2 + \dot{\tilde \sigma}^2 \right) \,.
}
The denominator of $\Omega$ is similarly related to the time derivative of the Hubble scale as
\eq{
\Omega = - 2 \frac{\dot{\tilde \sigma}\ddot{\tilde \theta}-\dot{\tilde \theta}\ddot{\tilde \sigma}}{\dot H} \,,
}
Now, in the slow--roll approximation, the second derivative with respect to time of the scalar fields shall be related to the gradients of the scalar potential. This is
\eq{
\ddot \phi_1^a &= - \frac{1}{3} \left( \frac{\dot V^a}{H} - \frac{V^a \dot H}{H^2} \right) \,,
}
with $V^a=\partial_a V$. We realize that the second term is suppressed by the slow--roll parameter as
\eq{
\ddot \phi_1^a &= - \frac{1}{3} \left( \frac{\dot V^a}{H} + \epsilon V^a  \right)
}
then, the numerator of $\Omega$ can be written as
\eq{
\dot{\tilde \sigma}\ddot{\tilde \theta}-\dot{\tilde \theta}\ddot{\tilde \sigma} = \frac{1}{3^2} \left( \frac{\dot V^1 V^2 - \dot V^2 V^1 }{H^2}  \right) 
}
where the upper index, indicates derivative with respect to the scalar fields $\{ 1,2 \} = \{ \tilde \theta , \tilde \sigma \}$. Let us define the vectors $\hat V = \langle V^1 , V^2 \rangle $ and $\hat V_O = \langle V^2 , -V^1 \rangle$ (notice that $\hat V \cdot \hat V_O = 0$), thus
\eq{
\dot{\tilde \sigma}\ddot{\tilde \theta}-\dot{\tilde \theta}\ddot{\tilde \sigma} = \frac{1}{3^2} \left( \frac{ \hat V^a \cdot d_t \hat V^a }{H^2}  \right) 
}
the $\Omega$ parameter is written as
\eq{
\Omega = \frac{2}{3^2} \left( \frac{ \hat V_O^a \cdot d_t \hat V^a }{\dot H H^2}  \right)
}
where in order to simplify notation we have used $d_t\hat{V}^a=\dot{\hat{V}}^a$. 
Thus in the slow--roll approximation, where $H^2 = \frac{1}{3} V$, the time derivative of the Hubble scale is given by
\eq{
 \dot H = \frac{1}{3^{1/2}\cdot 2} \frac{\dot V}{V^{1/2}} \,,
}
from which we get that
\eq{
\Omega = \frac{2^2}{3^{1/2}} \left( \frac{ \hat V_O^a \cdot d_t \hat V^a }{ \dot V V^{1/2}}  \right)
}
Thus, for constant fluxes any dependence of the form $V = \alpha F \left( \tilde \theta, \tilde \phi \right)$, with $\alpha$ a function on the fluxes, shall provide a parametric control of the fluxes over the angular velocity $\Omega$ of the form $\alpha^{1/2}$. However, for a scalar potential of such kind, the  ratio
\eq{
\frac{\Omega}{H} = 2^2 \left( \frac{ \hat V_O^a \cdot d_t \hat V^a }{ \dot V V}  \right) \,, 
}
cannot be parametrically controlled in terms of {\bf constant} fluxes. Notice that this is a generic result not exclusive to the specific model we are studying here. Therefore, one way to circumvent it is to consider  non-constant fluxes  \cite{Damian:2016lvj} which would give rise to terms of the form $\alpha^a$. This points out a relationship between the presence of inflationary directions and the possibility to control the ratio $\Omega/H$ by non-constant fluxes, and therefore, a way to have a numerical or parametrical control by fluxes to fulfill the swampland criterion.\\

\section{Final Remarks}

Based on the flux-scaling scenario we have constructed a simple two-field inflationary model in which besides the hierarchy of masses of the involved scales as well as the presence of inflationary trajectories, stable vacua and moduli stabilization at tree--level, we also have numerical control by fluxes on cosmological parameters as the spectral index, scalar-to-tensor ratio and non--Gaussianities. Moreover we show that by the incursion of non-geometric fluxes it is possible to uplift the stable non--supersymmetric AdS minimum to Minkowski or de Sitter. The inflationary trajectories correspond to those generated by uplifting flat directions by the presence of $p$-fluxes, which in turn breaks down the symmetry of the superpotential on the complex moduli $S$ and $T$.  However, as recently proposed, it is important to see wether inflation can coexist with an inflationary swampland criterion which establishes that effective inflationary models might be consistent to a theory of quantum gravity if the ratio $\Omega/H$ is of order one or larger. \\

For single-field scenarios this constraint rules out inflation, but not for multi-field inflation since $\Omega$ parametrizes the departure of inflationary trajectories to geodesics in the moduli space. We have shown that for our model this is indeed the case where  two-field inflation and the above mentioned swampland criterion can in fact coexist. It is important to mention that the inflaton corresponds to a linear combination of axions, which in the context of  F--theory inflation monodromy implies the absence of possible quantum corrections to the scalar potential.\\

A drawback in our approach is reflected on the impossibility to parametrically control the ratio $\Omega/H$ by fluxes. This follows from the approximation we are considering by selecting small values of $p$--fluxes.  One way to circumvent this result is to consider instead non--constant fluxes, e.g. fluxes depending on moduli.\\

However, as followed from the last assertion,  there is a subtle issue that must be stressed out. The existence of hierarchies, inflationary trajectories, and their subsequent parametrical control  is based on the assumption of non--integer values for the fluxes. Therefore, this model could be considered an effective realistic model if there is a way to enforce the fluxes to violate Dirac quantization. Mirror symmetry contributions can be responsible of this, but we leave such study for future work.\\

Finally, we want to comment about the dS vacuum constructed in our model and its obvious implications on the swampland bound. As stressed out in \cite{Obied:2018sgi, Agrawal:2018own}, the bound suggests that scalar potentials compatible to quantum gravity, as string theory,  have not stable dS vacua. In our case, the presence of non-geometric fluxes limits our model since there is not a complete compactification mechanism from a high energy theory from which the non-geometric fluxes appear. However, the flux-scale scenario we have followed, suggests that we can transfer such a role to the fluxes, meaning that non-constant and non-integer fluxes  could be the key to decouple quantum gravity effects on an effective potential, allowing in such limit to have inflation and dS vacua. Certainly a deeper study is required to have a better understanding about the relation between non-constant fluxes and the swampland criteria.

\vspace{1.0cm}

\begin{center}
{\bf Acknowledgments}
\end{center}

We thank Nana Cabo, Anamar\'ia Font, Xin Gao, Daniela Hershmann, Dami\'an Mayorga, Pramod Shukla,  Rui Sun,  and  Ivonne Zavala for useful discussions. We specially thanks Ralph Blumenhagen for explanations and suggestions. The work of C.D. is supported by a DAIP-Universidad de Guanajuato project No. CIIC 233/2018. O. L.-B. is supported by CONACyT project No. 258982 and by DAIP-Universidad de Guanajuato project No. CIIC 230/2018.

\newpage

\appendix

\section{\label{sec:level1}  Cosmological Observables and Slow-Roll Multi-field Inflation}

In the multifield scenario, for a homogenous and isotropic space-time the scalar fields shall satisfy the Einstein-Friedman equations, which in the N-fold formalism are given by
\eq{
\label{multiFriedman1}
H^2=\frac{1}{3\rm M_{pl}^2}\left[V(\phi^i)+\frac{1}{2} H^2 \cg_{ij} \frac{\partial \phi^i}{\partial N} \frac{\partial \phi^j}{\partial N} \right],
}
and
\eq{
\label{scalarfieldmulti}
\frac{d^2 \phi^i}{d N^2}+\left( 3 + \frac{1}{H} \frac{d H}{d N} \right)\frac{d \phi^i}{d N}+\frac{1}{H^2} \cg^{ij} \partial_j V=0,
}
where $\phi^i$ represent the i-th field, $dN = H dt$ and $\partial_i V=\partial V/\partial\varphi_i$. In the present work we shall specialize in a constant non-canonical kinetic terms for the scalar fields determined by the metric $\cg_{ab}$. This condition is obtained since the lightest fields are a linear combination of axions. As usual, inflation requires that the parameters%
\eq{ \label{epsilon}
\epsilon = - \frac{\dot H}{H^2} \,,
}
\eq{ \label{eta}
\eta = \frac{\dot \epsilon}{\epsilon H} \,,
}
shall satisfy $\eta \ll 1$ and $\epsilon \ll 1$ during inflation.  Thus, the number of e-folds is determined by
\eq{
\label{e-foldsmulti}
dN_e = H dt \,,
}
where its duration shall be around 40--60 in order to reproduce the observable universe. The explicit form of cosmological observables in the E-fold formalism are related to the scalar power spectrum \cite{Gao:2014fva} given by
\eq{ \label{ps}
P_{\mathcal{S}}^{(\chi,\zeta)}=\frac{H^2}{4\pi^2}\left( A^{\ca \cb} N_\ca N_\cb \right)_{\phi= \phi (e)},
}
where $N$ is given by the set of differential equations%
\eq{
\frac{d N_A}{d t} = - N_\cb P^\cb_\ca \,,
}
where the coeffients $P^{\cb}_\ca$ depends on the scalar potential as well to the time derivative of the scalar field as
\eq{
P^{\cb1}_{1\ca} = -\frac{1}{6 H^3} \left(  
   \begin{matrix} % or pmatrix or bmatrix or Bmatrix or ...
      \dth \p_\th V& \dth \p_\si V \\
      \dsi \p_\th V & \dsi \p_\si V \\
   \end{matrix} \right) \,,
}
\eq{
P^{\cb1}_{2\ca} = \frac{1}{6 H^3} 
   \left(  
   \begin{matrix} % or pmatrix or bmatrix or Bmatrix or ...
      (\p_\th V)^2& (\p_\si V)(\p_\th V) \\
      (\p_\si V)(\p_\th V) & (\p_\si V)^2 \\
   \end{matrix} 
   \right) \\ - 
   \frac{1}{6 H} 
   \left(  
   \begin{matrix} % or pmatrix or bmatrix or Bmatrix or ...
      \p^2_\th V		& \p_{\th \si} V \\
      \p_{\th \si} V 	& \p^2_\si V \\
   \end{matrix} 
   \right) \,,
   }
\eq{
P^{\cb2}_{1\ca} = -\frac{1}{6 H^3} 
   \left(  
   \begin{matrix} % or pmatrix or bmatrix or Bmatrix or ...
      \dth^2 & \dth \dsi \\
      \dth \dsi  & \dsi^2 \\
   \end{matrix} 
   \right)  - 
   \frac{1}{H} 
   \left(  
   \begin{matrix} % or pmatrix or bmatrix or Bmatrix or ...
     1	& 0 \\
      0 &1 \\
   \end{matrix} 
   \right) \,,
   }

\eq{
P^{\cb2}_{1\ca} = \frac{1}{6 H^3} 
   \left(  
   \begin{matrix} % or pmatrix or bmatrix or Bmatrix or ...
      \dth \p_\th V  & \dsi \p_\th V\\
      \dth \p_\si V &  \dsi \p_\si V \\
   \end{matrix} 
   \right)  - 
   3 
   \left(  
   \begin{matrix} % or pmatrix or bmatrix or Bmatrix or ...
     1	& 0 \\
      0 &1 \\
   \end{matrix} 
   \right) \,,
   }   
subject to the final condition 
\eq{
N_\ca ^F = - \left( \frac{H_\ca}{H_\cb F^\cb} \right) \,.
}
where $F^\cb_1 = \frac{1}{H} \dot \phi^\cb$, $F^\cb_2 = - 3 F^\cb_1- \frac{1}{H}\p^\cb V$, $H^1_\ca = \frac{1}{6 H} \p_\ca V$ and $H^2_\ca = \frac{1}{6H} \dot \phi_\ca$ . Thus, the spectral index is determined by the standard expression
\eq{
\label{ns}
n_{S}^{(\chi,\zeta)}-1=-2\epsilon-2\frac{A^{\ca \cb} N_\ca P^\cc_\cb N_\cc}{A^{\ca \cb} N_\ca N_\cb}+\frac{d_N A^{\ca \cb} N_\ca N_\cb}{A^{\ca \cb}N_\ca N_\cb} \,.
}
where $A^{\ca \cb}$ is calculated as shown in the Appendix of \cite{Gao:2014fva}. Similarly the tensor to scalar ratio $r$ is given by
\eq{
\label{r}
r^{(\chi,\zeta)}=8 \frac{ 1-(1+\alpha)\epsilon}{ N^\ca N_\ca }\,,
}
where $N^\ca = A^{\ca \cb} N_{\cb}$ and $\alpha=0.7296$. \\

Non-Gaussianities coming from the three point correlation functions related to the bi-spectrum are parametrized by $f_{NL}$.  Using the N-fold formalism  \cite{Maldacena:2002vr,Bernardeau:2002jy,Senatore:2010wk} which relates the curvature perturbations to the difference of e-foldings of two constant time hyper-surfaces, we have that
\eq{
\delta \phi^{\ca} \left( \lambda , N_e \right) = \phi^{\ca}  \left( \lambda + \delta \lambda , N_e \right) - \phi^\ca \left(  \lambda , N_e \right)
}
where $\lambda$ is an integration constant and $\phi^\ca$ stands for a scalar field. Thus, the curvature perturbations can be expressed as variations of the number of e-foldings. In particular, the $ f_{NL}$ parameters is expressed generically as%
\eq{
\label{fnl}
f_{\rm NL} =\frac{5}{6}\frac{N^\ca N^\cb N_{\ca \cb}}{\left( N^\cc N_\cc \right)^2},
}
where $N_{\ca \cb}$ is the solution of the differential equations
\eq{
\frac{d N_{\ca \cb}}{dt} = -N_{\ca \cc} P^{\cc}_\cb - N_{\cb \cc} P^{\cc}_\ca - N_\cc Q^\cc_{\ca \cb} \,,
}
where
\eq{
Q^{1 1 1}_{1 1 1} &=  \frac{1}{12 H^5} \dth (\p_\th V)^2 - \frac{1}{H^3} \dth (\p^2_{\th \th} V) \,, \\
Q^{1 1 1}_{1 2 1} & = Q^{1 1 1}_{1 1 2} =  \frac{1}{12 H^5} \dth (\p_\th V) (\p_\dsi V) - \frac{1}{H^3} \dth (\p^2_{\th \si} V) \,, \\
Q^{1 1 1}_{1 2 2} &= \frac{1}{12 H^5} \dth (\p_\si V)^2 - \frac{1}{H^3} \dth (\p^2_{\si \si} V) \,, \\
Q^{2 1 1}_{1 1 1} &= \frac{1}{12 H^5} \dsi (\p_\th V)^2 - \frac{1}{H^3} \dsi (\p^2_{\th \th} V) \,, \\
Q^{2 1 1}_{1 1 2} &=  Q^{2 1 1}_{1 2 1} =\frac{1}{12 H^5} \dsi (\p_\th V) (\p_\si V) - \frac{1}{H^3} \dsi (\p^2_{\th \si} V) \,, \\
Q^{2 1 1}_{1 2 2} &= \frac{1}{12 H^5} \dsi (\p_\si V)^2 - \frac{1}{H^3} \si (\p^2_{\si \si} V) \,, \\
}
\eq{
Q^{1 1 2}_{1 1 1} &= Q^{1 2 1}_{1 1 1} =   \frac{1}{12 H^5} \dth^2 (\p_\th V) - \frac{1}{H^3}  (\p_\th V) \,, \\
Q^{1 1 2}_{1 1 2} &= Q^{2 1 2}_{1 1 1} = Q^{2 2 1}_{1 1 1} =  Q^{1 2 1}_{1 2 1} =  \frac{1}{12 H^5} \dth \dsi \p_{\th} V \,, \\
Q^{1 1 2}_{1 2 1} &= Q^{1 2 1}_{1 1 2} =  \frac{1}{12 H^5} \dth^2 (\p_\si V) - \frac{1}{H^3}  (\p_\si V) \,, \\
Q^{2 1 2}_{1 1 2} &= Q^{2 2 1}_{1 2 1} = \frac{1}{12 H^5} \dsi^2 (\p_\th V) - \frac{1}{H^3}  (\p_\th V) \,, \\
Q^{2 1 2}_{1 2 1} &= Q^{1 1 2}_{1 2 2} = Q^{1 2 1}_{1 2 2} = Q^{2 2 1}_{1 1 2} = \frac{1}{12 H^5} \dth \si \p_{\si} V \,,\\
Q^{2 1 2}_{1 2 2} &= Q^{2 2 1}_{2 2 2} = \frac{1}{12 H^5} \dsi^2 (\p_\si V) - \frac{1}{H^3}  (\p_\si V) \,, \\
}
\eq{
Q^{1 2 2}_{1 1 1} = -\frac{3 \dth}{H^3}  + \frac{\dth^3 }{12 H^5} \,, Q^{1 2 2}_{1 1 2} = -\frac{\dsi}{H^3}  + \frac{\dth^2 \dsi}{12 H^5}  \,,\\
Q^{1 2 2}_{1 2 1} =-\frac{\dsi}{H^3}  + \frac{\dth \dsi^2}{12 H^5}  \,, Q^{1 2 2}_{1 2 2} =-\frac{\dth}{H^3}  + \frac{\dsi^3}{12 H^5}  \,,\\
Q^{2 2 2}_{1 1 1} =-\frac{3 \dsi}{H^3}  + \frac{\dth^3}{12 H^5}  \,, Q^{2 2 2}_{1 1 2} =-\frac{3\dth}{H^3}  + \frac{\dth^2 \dsi }{12 H^5} \,,\\
Q^{2 2 2}_{1 2 1} =-\frac{3 \dth}{H^3}  + \frac{ \dth \dsi^2}{12 H^5} \,, Q^{2 2 2}_{1 2 2} =-\frac{3\dsi}{H^3}  + \frac{\dsi^3}{12 H^5}   \,, \\
}
\eq{
Q^{1 1 1}_{2 1 1} &= -\frac{(\p_{\th} V)^3}{12 H^5} + \frac{\p_{\th} V \p^2_{\th \th} V}{2 H^3} - \frac{\p^3_{\th \th \th}V}{H} \,,\\
Q^{1 1 1}_{2 1 2} &=  -\frac{(\p_{\th} V)^2 (\p_{\si} V)}{12 H^5} + \frac{\p_{\si} V \p^2_{\th \th} V + 2 \p_\th V \p^2_{\th \si}V}{6 H^3} - \frac{\p^3_{\th \th \th}V}{H} \,, \\
Q^{1 1 1}_{2 2 1} &= -\frac{(\p_{\th} V)^2 (\p_{\si} V)}{12 H^5} + \frac{\p_{\si} V \p^2_{\th \th} V + 2 \p_\th V \p^2_{\th \si}V}{6 H^3} - \frac{\p^3_{\th \th \si}V}{H} \,, \\
Q^{1 1 1}_{2 2 2} &=  -\frac{(\p_{\th} V) (\p_{\si} V)^2}{12 H^5} + \frac{\p_{\th} V \p^2_{\si \si} V + 2 \p_\si V \p^2_{\th \si}V}{6 H^3} - \frac{\p^3_{\th \th \si}V}{H} \,, \\
Q^{2 1 1}_{2 1 1} &=  -\frac{(\p_{\th} V)^2 (\p_{\si} V)}{12 H^5} + \frac{\p_{\si} V \p^2_{\th \th} V + 2 \p_\th V \p^2_{\th \si}V}{6 H^3} - \frac{\p^3_{\th \si \si}V}{H} \,, \\
Q^{2 1 1}_{2 1 2} &= -\frac{(\p_{\th} V) (\p_{\si} V)^2}{12 H^5} + \frac{\p_{\th} V \p^2_{\si \si} V + 2 \p_\si V \p^2_{\th \si}V}{6 H^3} - \frac{\p^3_{\th \si \si}V}{H} \,, \\
Q^{2 1 1}_{2 2 1} &= -\frac{(\p_{\th} V) (\p_{\si} V)^2}{12 H^5} + \frac{\p_{\th} V \p^2_{\si \si} V + 2 \p_\si V \p^2_{\th \si}V}{6 H^3} - \frac{\p^3_{\si \si \si}V}{H} \,, \\
Q^{2 1 1}_{2 2 2} &=  -\frac{(\p_{\si} V)^3}{12 H^5} + \frac{\p_{\si} V \p^2_{\si \si} V}{2 H^3} - \frac{\p^3_{\si \si \si}V}{H} \,,\\
}
\eq{
Q^{1 1 2}_{2 1 1} &=  -\frac{\dth  (\p_\th V)^2}{12 H^5}  + \frac{\dth (\p^2_{\th \th} V)}{6 H^3}   \,, \\
Q^{1 1 2}_{2 1 2} &=  -\frac{\dsi  (\p_\th V)^2}{12 H^5}  + \frac{\dsi (\p^2_{\th \th} V)}{6 H^3}  \,,\\ 
Q^{1 1 2}_{2 2 1} &=  Q^{2 1 2}_{2 1 1} = -\frac{\dth  (\p_\th V) (\p_\si V)}{12 H^5}  + \frac{\dth (\p^2_{\si \th} V)}{6 H^3}  \,, \\
Q^{1 1 2}_{2 2 2} &= Q^{2 1 2}_{2 1 2} = -\frac{\dsi  (\p_\th V) (\p_\si V)}{12 H^5}  + \frac{\dsi (\p^2_{\si \th} V)}{6 H^3}  \,, \\
Q^{2 1 2}_{2 2 1} &= -\frac{\dth  (\p_\si V)^2}{12 H^5}  + \frac{\dth (\p^2_{\si \si} V)}{6 H^3} \, \\
Q^{2 1 2}_{2 2 2} &= -\frac{\dsi  (\p_\si V)^2}{12 H^5}  + \frac{\dsi (\p^2_{\si \si} V)}{6 H^3}  \,, \\
}
\eq{
Q^{1 2 1}_{2 1 1} &= -\frac{\dth  (\p_\th V)}{12 H^5}  + \frac{\dth (\p^2_{\th \th} V)}{6 H^3} \,, \\
Q^{1 2 1}_{2 1 2} &=  -\frac{\dth  (\p_\th V)}{12 H^5}  + \frac{\dth (\p^2_{\th \si} V)}{6 H^3} \,, \\
Q^{1 2 1}_{2 2 1} &=  -\frac{\dsi  (\p_\th V)}{12 H^5}  + \frac{\dsi (\p^2_{\th \th} V)}{6 H^3} \,, \\
Q^{1 2 1}_{2 2 2} &=  -\frac{\dsi  (\p_\th V)}{12 H^5}  + \frac{\dsi (\p^2_{\th \si} V)}{6 H^3} \,, \\
Q^{2 2 1}_{2 1 1} &= -\frac{\dth  (\p_\si V)}{12 H^5}  + \frac{\dth (\p^2_{\th \si} V)}{6 H^3} \,, \\
Q^{2 2 1}_{2 1 2} &= -\frac{\dth  (\p_\si V)}{12 H^5}  + \frac{\dth (\p^2_{\si \si} V)}{6 H^3} \,, \\
Q^{2 2 1}_{2 2 1} &= -\frac{\dsi  (\p_\si V)}{12 H^5}  + \frac{\dsi (\p^2_{\th \si} V)}{6 H^3} \,, \\
Q^{2 2 1}_{2 2 2} &= -\frac{\dsi  (\p_\si V)}{12 H^5}  + \frac{\dsi (\p^2_{\si \si} V)}{6 H^3} \,, \\
}
\eq{
Q^{1 2 1}_{2 1 1} &= -\frac{\dth^2 (\p_\th V)}{12 H^5}  + \frac{(\p_{\th} V)}{6 H^3} \,, \\
Q^{1 2 1}_{2 1 2} &= Q^{1 2 1}_{2 2 1} = -\frac{\dth \dsi \p_\th V}{12 H^5} \,, \\
Q^{1 2 1}_{2 2 2} &=-\frac{\dsi^2 (\p_\th V)}{12 H^5}  + \frac{ (\p_{\th} V)}{6 H^3} \,, \\
Q^{2 2 1}_{2 1 1} &=-\frac{\dth^2 (\p_\si V)}{12 H^5}  + \frac{ (\p_{\si} V)}{6 H^3} \,, \\
Q^{2 2 1}_{2 1 2} &= Q^{2 2 1}_{2 2 1} = -\frac{\dth \dsi \p_\si V}{12 H^5} \,, \\
Q^{2 2 1}_{2 2 2} &=-\frac{\dsi^2 (\p_\si V)}{12 H^5}  + \frac{ (\p_{\si} V)}{6 H^3} \,, \\
}
subject to the final condition
\eq{
N_{\ca \cb}^F = - \left( \frac{U_{\ca \cb}}{H_\cc F^{\cc}} \right) _{\phi= \phi (e)}.
}
Thus for two scalar fields we require to solve 16 differential equations.The explicitly form of $A^{\ca \cb}, P_\ca^\cb, U_{\ca \cb}, H_\ca$ and $F^{\ca}$ can be found in \cite{Gao:2014fva}.\\
%%% Use the following two code lines if you wish to generate your bibliography with BibTeX;
%%% please replace first the string "demo" below with the name(s) of
%%% the BibTeX data base(s) you want to use.
%%% The resulting bibliography-output (the contents of the .bbl file)
%%% must be pasted into this file before submission.
%%% 

\bibliographystyle{hep}
\bibliography{References}

\end{document}